\documentclass[3p,times,sort&compress,12pt]{elsarticle}

\usepackage{amssymb}

\usepackage{hyperref}

\usepackage{amsmath}
\usepackage{graphicx}
\usepackage{dcolumn}
\usepackage{bm}

\usepackage{threeparttable}

\usepackage{siunitx} 
\usepackage{physics} 
\usepackage{stmaryrd} 
\usepackage{mathtools}
\usepackage{txfonts}
\usepackage{subcaption}

\usepackage{xspace}
\makeatletter
\DeclareRobustCommand\onedot{\futurelet\@let@token\@onedot}
\def\@onedot{\ifx\@let@token.\else.\null\fi\xspace}

\def\ie{\emph{i.e}\onedot}

\makeatother

\usepackage[inline]{enumitem} %

\usepackage{soul,color} 
\usepackage[colorinlistoftodos,textsize=small]{todonotes}

\begin{document}

\begin{frontmatter}



\title{Simulation of high-speed impact of surfactant-laden drops}


\author[inst1]{Olivier D.~Y.~Huet}
\author[inst5]{Ravindra Pethiyagoda}
\author[inst1]{Timothy J.~Moroney}
\author[inst2,inst3]{Arvind Kumar}
\author[inst4]{Philip Taylor}
\author[inst2,inst3]{Justin J.~Cooper-White}
\author[inst1]{Scott W.~McCue \corref{cor}}
\cortext[cor]{Corresponding author}
\ead{scott.mccue@qut.edu.au}

\affiliation[inst1]{
            organization={School of Mathematical Sciences, Queensland University of Technology},
            city={Brisbane QLD 4000},
            country={Australia}
            }

\affiliation[inst5]{
            organization={School of Information and Physical Sciences, University of Newcastle},
            city={Newcastle NSW 2308},
            country={Australia}
            }

\affiliation[inst2]{
            organization={School of Chemical Engineering, University of Queensland},
            city={Brisbane QLD 4072},
            country={Australia}
            }

\affiliation[inst3]{
            organization={Australian Institute for Bioengineering and Nanotechnology, University of Queensland},
            city={Brisbane QLD 4072},
            country={Australia}
            }

\affiliation[inst4]{
            organization={Syngenta},
            addressline={Jealott’s Hill International Research Centre},
            postcode={Bracknell RG42 6EY},
            country={United Kingdom}
            }

\begin{abstract}
We develop a computational model to simulate the immediate post-impact spreading behaviour of surfactant-laden drops that impact a flat and solid surface.  The model is built on the InterFoam solver (OpenFOAM software), which uses the volume-of-fluid method to solve the Navier-Stokes equations. In order to incorporate surfactant in the bulk and on the interface, we make numerous modifications and extensions, such as coupling the volume-of-fluid method with a level-set method. Simulations demonstrate the accumulation of surfactant in the vicinity  of the moving contact line, especially during the formation of the rim. Gradients of surfactant at the liquid-air interface lead to Marangoni forces that oppose the drop spreading, while high-velocity impacts reduce the overall surface tension and increase the magnitude of Marangoni forces. Both of these phenomena, which tend to reduce the maximum spreading, are highly dependent on the surfactant properties. Our computational methodology expands the potential for utilising Computational Fluid Dynamics to model complex interfacial flows that involve surfactants, leading to various opportunities in the future.
\end{abstract}




\end{frontmatter}


\section{Introduction}

Liquid drops impacting on solid surfaces have garnered significant attention from researchers and industries alike due to wide-ranging applications, such as inkjet printing \citep{derby2010inkjet} and pesticide application \citep{dorr2014towards,dorr2016spray}, where the deposition of drop onto surfaces is a critical step.
Although extensive experimental research has been conducted on the impact and wetting of pure liquids on solid surfaces \citep{bonn2009wetting,josserand2016drop,rioboo2001outcomes,rioboo2002time,fang2022drop}, more studies are currently ongoing to explore the behaviour of drops that include additives. Additives such as polymers or particles can improve liquid retention during impact \citep{crooks2001role,cooper2002dynamics,grishaev2015complex}. In this computational study, we focus on the addition of surfactants (surface-active agents) that promote drop retention due to its ability to reduce surface tension \citep{massinon2017spray,gao2020wetting,esmaeili2021further}.

Surfactant sorption is a complex and dynamic phenomenon encompassing the adsorption and desorption of surfactant molecules at the interfaces between liquid-gas or liquid-solid. The quantity of surfactant that can be adsorbed at these interfaces and the speed at which surfactants are transferred in or out of them depend on many properties, such as molecular weight, shape, and concentration of the surfactant but also physicochemical properties of the solvent \citep{landoll1982nonionic}.
The complexity of sorption increases further during dynamic events where the liquid is in motion and the interface undergoes deformation. For instance, during an impact event, the distribution of surfactants at the liquid-gas interface is disrupted, leading to the rise of Marangoni forces due to the creation of a surfactant concentration gradient along the interface \citep{zhang1997dynamic}. Surfactants also modify the drop impact dynamics by affecting the contact angle \citep{wang2022unconditionally,varghese2024effect}.

Experiments serve as a direct means of investigating the effect of surfactants on drop impaction, offering valuable insights into this phenomenon. On the other hand, numerical simulations present a flexible and cost-effective alternative, capable of providing unique insights into the underlying mechanisms governing drop impact, which are challenging to obtain experimentally. However, Computational Fluid Dynamics (CFD) requires a comprehensive understanding of the underlying physics and chemistry to accurately simulate the process.


To the best of our knowledge, among the very few published studies using CFD methodology to simulate the behaviour of surfactant-laden droplets spreading or receding on solid substrates,
most restrict themselves to liquids containing a low concentration of soluble surfactant with no vertical impact (and therefore negligible initial inertia) \citep{badra2018numerical,antritter2019spreading,gao2021surfactant,wang2022unconditionally}, while \cite{ganesan2015simulations} and \cite{wang2022energy} are concerned with drop impaction driven by inertia, this time with moderate concentration of surfactant.
Our paper distinguishes itself by focusing on scenarios where:
\begin{enumerate*}[label=(\roman*)]
\item inertia serves as the primary driver for the initial liquid-air interface deformation,
\item the initial surfactant concentration is relatively high (albeit below the critical micelle concentration), and
\item the interface of the drop is at equilibrium prior to impact.
\end{enumerate*}
Our approach differs from \cite{ganesan2015simulations} and \cite{wang2022energy} in that we implement the volume-of-fluid (VOF) method, whereas they use schemes based on the finite-element and phase-field methods, respectively.  Note that in this preliminary study, we restrict our simulations to an axisymmetric geometry where the drop impacts the substrate surface perpendicularly. The case in which the drop shatters (such as in \citet{bussmann2000modeling}, e.g.) is not considered here.

The computational framework employed in this study is described in detail in Section~\ref{IJMF2023:sec:compWork}. Our model is based on the VOF method, which solves the Navier-Stokes equations for a mixture of liquid and air without explicitly imposing boundary conditions at the liquid-air interface \citep{gopala2008volume}, and has proven to be a popular framework for simulating drop impaction on solid substrates \citep{pasandideh1996capillary,cimpeanu2018three,debnath2023understanding,henman2023computational,hu20223d}.  In order to accurately consider the effects of soluble surfactants on interfaces, several significant modifications to the standard approach are necessary. In the first class of modifications, we introduce a modified coupled level-set VOF method for fluid advection, which is detailed in Section~\ref{IJMF2023:sec:compWorkCLSVOF}. Secondly, a modified surfactant model is presented in Section~\ref{IJMF2023:sec:compWorkSaa}. Simulations are presented in Section~\ref{IJMF2023:sec:results} to demonstrate the effectiveness of our newly developed model and gain insights into the dynamics of surfactant-laden drops. Finally, our study is concluded in Section~\ref{IJMF2023:sec:ccl}.

This study highlights many of the considerable computational challenges involved when simulating drop impaction with surfactant and provides details of how these challenges can be overcome within a VOF context. Our work opens up many future possibilities for using CFD to simulate interfacial flows with soluble surfactants, involving drop impaction and other applications. The entirety of our code is implemented within the software OpenFOAM, which is freely available and open-source; it is hosted on the first author's personal \href{https://github.com/OlivierDY-Huet/surfactantInterFoamPlus.git}{Github page}.




\section{Problem formulation}

The purpose of the model developed in this study is to investigate the impact behaviour of surfactant-laden drops on a solid flat surface, specifically focusing on moderately high impact velocities and initial surfactant concentrations that approach the critical micelle concentration. The drops consist of a solvent liquid, with specific properties including surface tension $\sigma_0$, density $\rho_\ell$, and dynamic viscosity $\mu_\ell$.
Immediately before impact, the drop is modelled as a sphere with a diameter $D_0$ falling at a speed $V_0$. The drop contains a concentration of surfactant $C_\textrm{B}^0$, and its interface concentration is in equilibrium leading to an equilibrium surface tension $\sigma_\textrm{eq}$.

Since the OpenFOAM software we build our model on operates using dimensional quantities, we maintain dimensional variables in our model to align with the software. To demonstrate the model's capabilities, we shall present simulations encompassing a range of Weber numbers, 
\begin{equation*} \label{IJMF2023:eqn:We}
\textrm{We}=\frac{\rho_\ell V_0^2 D_0}{\sigma_\textrm{eq}},
\end{equation*}
from approximately 0.35 to 10, and Reynolds numbers, 
\begin{equation*} \label{IJMF2023:eqn:Re}
\textrm{Re}=\frac{\rho_\ell V_0 D_0}{\mu_\ell},
\end{equation*}
spanning approximately 500 to 2000 prior to impact.

\begin{figure}[!htbp]
\centering\includegraphics[width=0.7\linewidth]{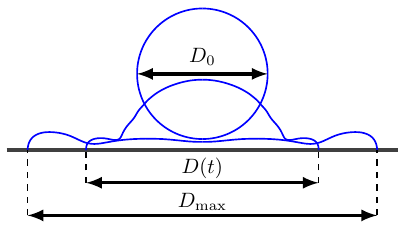}
\caption{Schematic of a drop undergoing deformation upon impact, which results in a change in the drop diameter $D(t)$ from its initial value $D_0$ to the maximum diameter $D_\textrm{max}$ at the end of the spreading phase.}
\label{IJMF2023:fig:1}
\end{figure}

During impact, the drop undergoes deformation, resulting in a change of the circular contact line's diameter $D(t)$ up to the maximum diameter $D_\textrm{max}$ that is achieved at the end of the spreading phase, as illustrated in Figure \ref{IJMF2023:fig:1}. The maximum relative diameter $\beta_\textrm{max} = D_\textrm{max}/D_0$ serves as a key measure to assess the overall influence of interfacial forces on impact dynamics. Since surfactants solely modify the properties of the liquid-gas interface and the contact line, this study concentrates on the quantity $\beta_\textrm{max}$ to study their effect on the impact dynamics.


\section{Computational framework} \label{IJMF2023:sec:compWork}

\subsection{Preamble} \label{IJMF2023:sec:preamble}

This study presents a new CFD model that simulates the impact of surfactant-laden drops on flat surfaces. Although the simulations are focused on this particular physical application, the computational framework described in this section could be applied to a wide range of CFD simulations involving interfacial flows with surfactants that interact with substrates. The model is based on the InterFoam solver of the OpenFOAM software \citep{OpenFOAM}, which employs the VOF method to accurately capture the interface between two immiscible, isothermal, incompressible fluids by solving the Navier Stokes equations. The modifications and extensions of InterFoam introduced in this study enable the transport of surfactants as well as the sorption phenomena that occur between the bulk, the liquid-air interface, and the substrate. During the impact, the interface deformation of the drop modifies the surfactant concentration at the liquid-gas interface, resulting in a surface tension gradient. The resulting Marangoni forces can affect the drop spreading dynamics, although the extent to which they contribute significantly can be difficult to appreciate.

While the model we present is compatible with a three-dimensional (3D) geometry, the resulting simulations are computationally expensive. To reduce computational expense, we employ a wedge domain for axisymmetric simulations. Therefore, the scope of this study is limited to drops that do not undergo shattering, as numerically accurate film-breaking necessitates a complete 3D domain. By employing the wedge domain, we leverage computational resources to attain higher-resolution simulations for axisymmetric cases.

Prior to describing the model, it is important to clarify that it is designed specifically for liquid mixtures whose initial surfactant concentrations fall below the critical micelle concentration (CMC), which is defined as the minimum concentration of surfactant required for the formation of micelles in a solvent. The reason why micelles are not included in the initial drop is twofold. Firstly, the model already represents significant improvement over existing models, even in the absence of micelles. Secondly, high concentrations of micelles lead to an acceleration of the adsorption process \citep{fainerman2006dynamic}. This effect can result in a drop with nearly constant surface tension at any given impact intensity. Although this phenomenon may be relevant in certain contexts, our investigation concentrates primarily on investigating the impact dynamics influenced by Marangoni effects, which arise from surface tension gradients. Therefore, this study focuses on cases where micelles are absent from the falling drop.

\subsection{Coupled level-set volume-of-fluid method} \label{IJMF2023:sec:compWorkCLSVOF}

The model in this study relies on the algebraic coupled level-set volume-of-fluid (A-CLSVOF) method to simulate multiphase flows \citep{haghshenas2017algebraic}. This numerical technique combines two well-established approaches, namely the volume-of-fluid (VOF) method and the level-set method (LS).

The VOF method is a technique that discretises the fluid domain into cells and assigns a volume fraction to each cell to represent the fraction of each fluid in that cell.
This method is established for its ability to accurately preserve mass during simulations. However, a weakness of the VOF method is the generation of spurious flows, which can be attributed to inaccuracies in the interface description.

The LS method is a technique that assigns a value to each cell to represent the distance between the center of the cell and the nearest point on the interface. In contrast to the VOF method, the LS method provides an accurate interface description but does not guarantee mass conservation.
In the A-CLSVOF method, the LS and VOF methods are combined to take advantage of their respective strengths. The volume fraction is used to calculate the fluid properties and to solve the Navier-Stokes equations. The LS function is updated based on the volume fraction and then used to calculate the interfacial forces that are required to advect the fluids.

\subsubsection{Volume-of-fluid method} \label{IJMF2023:sec:VOF}

When a liquid drop impacts a solid surface, it involves the interaction between two distinct fluid phases: liquid and air. To study this phenomenon, the VOF method is used, which models the two phases as a single fluid mixture \citep{gueyffier1999volume,scardovelli1999direct}.  The properties of this mixture depend on the volume fraction $\alpha$, which varies between 0 and 1: $\alpha=0$ represents pure air, $\alpha=1$ represents pure liquid, while values of $\alpha$ close to 0.5 are associated with the liquid-gas interface. The density $\rho$ and dynamic viscosity $\mu$ of the mixture are calculated for each point in space and time using
$\rho =\alpha \rho_{\ell} + (1-\alpha) \rho_{\textrm{g}}$
and
$\mu=\alpha \mu_{\ell} + (1-\alpha) \mu_{\textrm{g}}$,
where $\rho_{\ell}$ and $\mu_{\ell}$ denote the liquid properties, while $\rho_{\textrm{g}}$ and $\mu_{\textrm{g}}$ denote the air properties.

Using the VOF method, the motion of the mixture is modelled using the mass conservation equation (incompressible),
\begin{equation} \label{IJMF2023:eqn:incompressible}
\grad \cdot \boldsymbol{q}
=0,
\end{equation}
the momentum conservation equation (Navier-Stokes)
\begin{equation} \label{IJMF2023:eqn:NavierStokes}
\frac{\partial (\rho \boldsymbol{q})}{\partial t}
+ \grad \cdot \left( \rho \left(\boldsymbol{q} \otimes \boldsymbol{q} \right)\right)
=
\grad \cdot \boldsymbol{T}
-\grad p
+ \rho \boldsymbol{g}
+ \boldsymbol{f}_{\sigma},
\end{equation}
and the volume fraction advection equation
\begin{equation} \label{IJMF2023:eqn:alphaAdvOF}
\frac{\partial \alpha}{\partial t}
+ \grad \cdot \left( \alpha \boldsymbol{q} \right)
+ \grad \cdot \left( \alpha \left( 1 - \alpha \right) \boldsymbol{q}_{\textrm{r}} \right)
=0,
\end{equation}
where $\boldsymbol{q}$ is the velocity vector, $p$ is the pressure, $\boldsymbol{g}$ is the gravitational acceleration, $\boldsymbol{f}_{\sigma}$ are the forces located at the interface (more detail in Section~\ref{IJMF2023:sec:intfForces}),  $\boldsymbol{T} = \mu ( \nabla \boldsymbol{q} + (\nabla \boldsymbol{q})^T )$ is the viscous stress tensor, and $\boldsymbol{q}_{\textrm{r}} = c_{\textrm{r}} \norm{\boldsymbol{q}} \boldsymbol{n}$ is the relative compressive velocity. The vector field $\boldsymbol{n}$ provides the normal direction of the interface pointing inward, which is detailed in Section~\ref{IJMF2023:sec:LS}, while $c_{\textrm{r}}$ is the relative compression coefficient.

The relative compressive velocity limits the numerical diffusion at the interface \citep{okagaki2021numerical}. The relative compression coefficient is typically assigned a value of 1 in order to achieve optimal compression in the simplified scenario where the coefficient is assumed to remain constant \citep{deshpande2012evaluating}. In this study, $c_{\textrm{r}}$ is adaptive \citep{mehmani2018wrinkle}:
\begin{equation}
\label{IJMF2023:eqn:adaptComp}
c_{\textrm{r}}
=
\frac{\cos{\left(2 \arccos{\left| \boldsymbol{n} \cdot \boldsymbol{n}_f \right|}\right)}+1}{2}
,
\end{equation}
where $\boldsymbol{n}_f$ denotes the cell face normal. The adaptive compression coefficient varies between 0 and 1 to maintain a smooth (wrinkle-free) yet sharp interface.

\subsubsection{Level-set method} \label{IJMF2023:sec:LS}

The LS method outlined in this section follows the methodology outlined in \cite{haghshenas2017algebraic}. However, adjustments are necessary to tackle challenges that arise during drop impact due to the interplay between the drop interface and the substrate.

The LS method relies on a signed distance function, denoted as $\psi$, which takes positive values inside the liquid, negative values outside, and is zero at the liquid-air interface. The gradient of the LS field, $\grad \psi$, is a dimensionless vector field with a unit magnitude that points into the liquid. This property makes $\grad \psi$ an ideal choice for $\boldsymbol{n}$; therefore, the normal direction to the interface is defined globally as $\boldsymbol{n} = \grad \psi$. To ensure that $\norm{\grad \psi} = 1$ even in the vicinity of the substrate, we compute the gradient exclusively using the cell values while disregarding the boundary value. As a result, only the $\psi$ values  within the domain affect the magnitude of $\grad \psi$.

At each time step, $\psi$ is initialised with an initial distance field denoted as $\psi_{0}$, which is calculated via
\begin{equation*}
\label{IJMF2023:eqn:psi0}
\psi_{0}=(2\alpha-1)\frac{\varepsilon}{2},
\end{equation*}
where $\varepsilon = C_\varepsilon h$ represents the thickness of the interface, with $C_\varepsilon=1.4$, and $h$ denotes the grid size. The use of an adaptive compression coefficient compared to using a constant $c_{\textrm{r}}$, as presented in Section \ref{IJMF2023:sec:VOF}, yields a smoother $\alpha$, resulting in a smoother $\psi_0$ that is a more accurate approximation of $\psi$ near the interface. This modification in the VOF method improves the convergence stability of $\psi$ by improving $\psi_0$.

After initialising $\psi$ with $\psi_{0}$, the sign distance function undergoes modification according to the equation
\begin{equation}
\label{IJMF2023:eqn:psiReIni}
\frac{\partial \psi}{\partial \tau} + \lvert S(\psi) \left(G(\psi)-1 \right) \rvert S(\psi) \left(G(\psi)-1 \right)  = 0
\end{equation}
over an artificial interval time $\tau$, where $S(\psi)$ represents a smooth sign function, and $G(\psi)$ denotes the Hamilton-Godunov function, which is a specific method for evaluating $\norm{\grad \psi}$. Before delving into the details of calculating $S(\psi)$ and $G(\psi)$, we note that $S(\psi)$ is zero at the interface and $(G(\psi)-1)$ approaches zero as $\norm{\grad \psi}$ tends to one. Consequently, the values of $\psi$ at the interface are scarcely updated, preventing deviations of $\psi=0$ from $\alpha=0.5$. Conversely,  $\psi$ is more extensively updated in the remaining domain until $\norm{\grad \psi} \approx 1$.
In the original work by \cite{haghshenas2017algebraic}, the second term in
(\ref{IJMF2023:eqn:psiReIni}) is given by $S(\psi) (G(\psi)-1)$. However, we employ the modification $\lvert S(\psi) \left(G(\psi)-1 \right) \rvert S(\psi)  \left(G(\psi)-1 \right)$ instead, as it enhances the convergence rate of $\psi$ and limits the deviations of $\psi=0$ from $\alpha=0.5$ in the substrate vicinity. The LS field is more prone to deviations at the substrate due to the disruption of local symmetry in the distance function caused by the values of the contact angle away from 90$\si{\degree}$, particularly as the contact angle approaches 0$\si{\degree}$ or 180$\si{\degree}$.

The function $S$ is defined to be zero when $\psi_{0}=0$ or $\alpha=0.5$, ensuring that the position of the interface obtained from the LS method is consistent with that calculated from the VOF method. In addition to preventing the migration of the zero level set, the sign function enables $\psi$ to converge smoothly near the interface. The sign function in \cite{haghshenas2017algebraic} adopted from \cite{peng1999pde} is infinitely smooth.
In this study, we use
\begin{equation*}
\label{IJMF2023:eqn:signfct}
S(\psi) =
\begin{cases}
-1 & \psi < -\varepsilon \omega \\
\displaystyle \frac{\psi}{\varepsilon \omega} + \frac{1}{\pi} \sin{ \left( \frac{\pi \psi}{\varepsilon \omega} \right)}&  -\varepsilon \omega \leq \psi \leq \varepsilon \omega \\
1 &  \psi > \varepsilon \omega \\
\end{cases},
\end{equation*}
as proposed by \cite{mousavi2016level}, where $\omega = \min(\norm{\grad \psi},1)$. This definition of $S$ is not infinitely smooth but possesses a finite width. As a result, we speed up the convergence rate of $\psi$ to prevent the migration of $\psi=0$, although this comes at the expense of reduced convergence stability.

The Hamilton-Godunov function $G$, as described in \cite{bardi1991nonconvex}, relies on the upwind and downwind first derivatives of $\psi$ along the principal coordinate directions. This function accounts for the variations of $\psi$ that occur outward from the interface, rather than inward. Consequently, during the update of $\psi$, only the $\psi$ values closer to the interface are considered, resulting in improved accuracy of the update process.

After achieving convergence of the distance function $\psi$, an accurate normal vector $\boldsymbol{n}$ can be obtained. The numerical computation of the normal vector involves the calculation of the expression $\grad \psi \textrm{/} \norm{\grad \psi}$, ensuring the normalisation of the vector. Additional variables, including the smoothed Heaviside function $\mathcal{H}$ and the Dirac function $\mathcal{D}$, are derived from $\psi$ as follow:
\begin{equation*}
\label{IJMF2023:eqn:Dirac}
\mathcal{D}=
    \begin{cases}
    0
    &\textrm{if $\lvert \psi \rvert > \varepsilon$}\\
    \displaystyle \frac{1}{2 \varepsilon} \left( 1 + \cos{\left( \frac{\pi \psi}{\varepsilon}\right)} \right)
    &\textrm{if $\lvert \psi \rvert \leq \varepsilon$}
    \end{cases}
    ,
\end{equation*}
and
\begin{equation*}
\label{IJMF2023:eqn:Heaviside}
\mathcal{H}=
    \begin{cases}
    0
    &\textrm{if $\psi < -\varepsilon$}\\
    \displaystyle \frac{1}{2} \left( 1 + \frac{\psi}{\varepsilon} + \frac{1}{\pi} \sin{\left( \frac{\pi \psi}{\varepsilon}\right)} \right)
    &\textrm{if $-\varepsilon \leq \psi \leq \varepsilon$}\\
    1
    &\textrm{if $\psi > \varepsilon$}
    \end{cases}
    .
\end{equation*}

Using the A-CLSVOF method, the volume fraction advection equation presented in Section~\ref{IJMF2023:sec:VOF} becomes
\begin{equation} \label{IJMF2023:eqn:alphaAdvMod}
\frac{\partial \alpha}{\partial t}
+ \grad \cdot \left( \alpha \boldsymbol{q} \right)
+ \grad \cdot \left( \alpha \left( 1 - \mathcal{H} \right) \boldsymbol{q}_{\textrm{r}} \right)
=0.
\end{equation}
The incorporation of the level-set field within the third term enhances the artificial compression of the interface.

\subsubsection{Interfacial forces} \label{IJMF2023:sec:intfForces}

The delta function $\delta_{\textrm{I}}$, which measures the interface area per unit volume, is defined as $\norm{\grad \alpha}$ in the VOF method. However, in the LS method, $\delta_{\textrm{I}}$ is computed as either $\norm{\mathcal{D} \grad \psi}$ or $\norm{\grad \mathcal{H}}$, depending on the chosen implementation. Both LS definitions are mathematically equivalent but yield slightly different numerical results \citep{haghshenas2017algebraic}. For this study, we adopt the definition based on the Dirac function $\mathcal{D}$. The delta function $\delta_{\textrm{I}}=\norm{\mathcal{D} \grad \psi}$ is nonzero only in the vicinity of the interface and plays a crucial role in scaling the interfacial forces.

At the liquid-air interface, there are two main forces: surface tension and Marangoni forces. Surface tension forces are normal to the interface and arise from the cohesive forces between the molecules of the liquid and tend to minimise the surface area of the liquid. Marangoni forces arise from the gradient of surface tension along the liquid interface. This surface tension gradient generates a force along the interface. Therefore, the forces at the liquid-gas interface, denoted as $\boldsymbol{f}_{\sigma}$ in (\ref{IJMF2023:eqn:NavierStokes}), can be expressed as follows:
\begin{equation}
\label{IJMF2023:eqn:STF}
\boldsymbol{f}_{\sigma}
=
- \delta_{\textrm{I}}^\textrm{scaled} \kappa  \boldsymbol{n} \sigma
+ \delta_{\textrm{I}}^\textrm{scaled} (\varmathbb{I} - \boldsymbol{n} \boldsymbol{n}) \grad \sigma,
\end{equation}
where $\sigma$ is the surface tension coefficient, $\kappa=\div \boldsymbol{n}$ represents the interface curvature, $\varmathbb{I}$ is the identity tensor, and $\delta_{\textrm{I}}^\textrm{scaled} = 2 \mathcal{H} \delta_{\textrm{I}}$ scales both forces to act only at the interface. The scaling factor $2 \mathcal{H}$ accounts for the density jump across the interface in the level set method, ensuring that the acceleration is approximately symmetric across the interface and reducing spurious flows \citep{yokoi2014density,gu2019volume}. Note that Equation~(\ref{IJMF2023:eqn:STF}) contains $\sigma$, which is not constant in our model, but rather is dependent on the concentration of surfactants present at the liquid-gas interface.

\subsubsection{Boundary conditions}

In our model, there are two types of boundaries defining the computational domain. The first corresponds to the lower surface where the substrate is located, while the second represents the outer surface of the domain that does not encompass the substrate. To maintain the integrity of our simulation, it is crucial that the second type of boundary remains sufficiently distant from the impact point, so that $\alpha \approx 0$ there.

The velocity $\boldsymbol{q}$ and pressure $p$ at the boundaries are interdependent.
The pressure values at the domain boundaries are determined based on the local velocity and the constant total pressure within the domain.
The velocity follows a no-slip condition at the substrate while maintaining a zero gradient at the other domain boundaries.
Regarding the volume fraction $\alpha$, we adjust its gradient direction at the substrate to ensure that the contact angle matches the value specified in the Kistler model (refer to Section~\ref{IJMF2023:sec:Kistler}). As for the remaining domain boundaries, the boundary conditions for $\alpha$ are determined by the direction of the flow: we enforce a constant value of 0 for inflows (flows entering the domain) to prevent any liquid from entering the domain, while we apply a zero gradient condition for outflows (flows leaving the domain).

\subsection{Surfactant model} \label{IJMF2023:sec:compWorkSaa}

In this study, the model accounts for the influence of surfactants. When soluble surfactants are added to the solution, they are present not only in the bulk liquid phase but also at the liquid-gas interface where they reduce the surface tension due to their amphiphilic nature. During the impact of a surfactant-laden drop on a substrate, the surfactant concentration in both the bulk and at the interface changes dramatically due to the deformation of the interface and sorption of surfactants on the substrate. The alteration in surfactant concentration at the liquid-gas interface modifies the surface tension and interfacial forces, leading to changes in the drop impact dynamics.

Previous studies, such as those focusing on autophobing \citep{craster2007autophobing,bera2016surfactant} and superspreading  \citep{theodorakis2015superspreading,theodorakis2019molecular} phenomena, have demonstrated that the interaction between surfactants at the liquid-gas interface and the substrate can significantly influence the drop behaviour on the substrate.
%
%
To our knowledge, existing CFD frameworks employed to simulate the impact of surfactant-laden drops have not accounted for the substrate's influence on surfactant concentration. In contrast, our CFD framework incorporates this factor, enabling the substrate to change the drop's surfactant concentration and affect the drop impact dynamics.

\subsubsection{Surfactant governing equation}

The equation that governs the surfactant concentration in the bulk $c_{\textrm{B}}$ is
\begin{equation}
    \label{IJMF2023:eqn:c_B}
    \frac{\partial c_{\textrm{B}}}{\partial t}
    + \grad \cdot \left( c_{\textrm{B}} \boldsymbol{q} \right)
    + \grad \cdot \left( c_{\textrm{B}} \left( 1 - \mathcal{H} \right)\boldsymbol{q}_{\textrm{r}} \right)
    =
    D_{\textrm{B}} \grad^2 c_{\textrm{B}}
    - \grad \cdot \left( \frac{c_{\textrm{B}}}{\mathcal{H}} D_{\textrm{B}} \grad \mathcal{H} \right)
    + j_{\textrm{BI}} + j_{\textrm{BW}},
\end{equation}
where the total surfactant quantity in bulk can only be modified through the source-sink terms $j_{\textrm{BI}}$ and $j_{\textrm{BW}}$ that depend on the surfactant interactions between the bulk and the liquid-gas interface, and between the bulk and the substrate, respectively. The first and second terms on the right-hand side of the equation ensure that the diffusion of $c_{\textrm{B}}$ is limited to the bulk. The inclusion of the third term on the left-hand side prevents numerical diffusion of $c_{\textrm{B}}$, similar to the approach employed in (\ref{IJMF2023:eqn:alphaAdvOF}).

The temporal evolution of the surfactant concentration at the liquid-air interface $c_{\textrm{I}}$ is governed by
\begin{equation}
    \label{IJMF2023:eqn:c_I}
    \frac{\partial c_{\textrm{I}}}{\partial t}
    + \grad \cdot \left( c_{\textrm{I}} \boldsymbol{q} \right)
    + \grad \cdot \left( c_{\textrm{I}} \left( 1 - 2 \mathcal{H} \right)\boldsymbol{q}_{\textrm{r}} \right)
    =
    D_\lambda \grad_\Sigma^2 c_{\textrm{I}}
    + j_{\textrm{IB}} + j_{\textrm{IW}},
\end{equation}
where $D_\lambda$ is the diffusion coefficient along the interface, $j_{\textrm{IB}} (= - j_{\textrm{BI}})$  and $j_{\textrm{IW}}$ are the sink-source terms that occur, respectively, between the liquid-air interface and the bulk, and between liquid-air interface and the substrate. The special Laplace–Beltrami operator $\grad_\Sigma^2$ ensures that the diffusion of $c_{\textrm{I}}$ is tangential to the interface.  The third term on the left-hand side of (\ref{IJMF2023:eqn:c_I}) incorporates the factor $(1-2\mathcal{H})$, which causes the artificial velocity to be directed towards the interface.

The equation governing the concentration of surfactant on the substrate (referred to as ``wall'' in OpenFOAM) $c_{\textrm{W}}$ is expressed as
\begin{equation}
    \label{IJMF2023:eqn:c_W}
    \frac{\partial c_{\textrm{W}}}{\partial t} + \grad \cdot \left( c_{\textrm{W}} \boldsymbol{q}_{\textrm{W}} \right)
    =
    j_{\textrm{WI}} + j_{\textrm{WB}},
\end{equation}
where \textcolor{black}{$\boldsymbol{q}_{\textrm{W}}$ is the velocity vector at the substrate while}, $j_{\textrm{WI}} (=-j_{\textrm{IW}})$ and $j_{\textrm{WB}} (=-j_{\textrm{WB}})$ are the respective source-sink terms representing the flux of surfactant molecules towards the liquid-gas interface and the bulk.

\subsubsection{Simplifications and assumptions}

Simulating the behaviour of a surfactant-laden drop presents several challenges. To overcome some of these, it is necessary to make assumptions and simplifications, which help to remove interactions that are not yet well understood, avoid parameters that are difficult to measure, and reduce the computational cost, while retaining sufficient information to deepen our understanding of the drop impact dynamics.

Surfactants in a bulk solution can aggregate into micelles if the concentration surpasses the CMC. Surfactants that form micelles exhibit distinct properties from their isolated state, including a different adsorption kinetics \citep{danov2006mass}. The model presented in this study is tailored to drops whose initial surfactant concentration falls below the CMC, as stated in Section~\ref{IJMF2023:sec:preamble}. However, during drop impact dynamics, the surfactant concentration may exceed the CMC due to the desorption of the interface, resulting in the formation of micelles. It is assumed that the number of micelles remains small, thus having a negligible effect on the drop impact dynamics. Therefore, the surfactant concentration in our model accounts for both isolated and micellar forms, assuming that micelles share the same properties as the free form. This simplification was implemented to avoid introducing a new governing equation, which would significantly increase the complexity of the system of equations.

\textcolor{black}{Turning to the substrate, in our study we apply a non-slip boundary condition at the substrate ($q_\textrm{W} = 0$), instead of a Navier-slip condition. This simplification serves two purposes: firstly, it lessens the complexity of our already intricate model by not requiring an additional slip length parameter that is very difficult to approximate. Secondly, the advection term in equation (\ref{IJMF2023:eqn:c_W}) can be neglected, meaning the surfactant molecules remain immobile once adsorbed onto the substrate.}  To further simplify our model, we only consider the spreading phase of the drop impact and assume that the surfactant interaction between the substrate and the bulk can be neglected  ($j_{\textrm{WB}} = j_{\textrm{BW}} = 0$). This is because such interaction occurs away from the triple line and does not significantly affect the surface tension or dynamics of the impact during the spreading phase. However, if we were to simulate the receding phase, this interaction cannot be ignored since the contact line would interact with the substrate whose properties were modified during the first phase.

\textcolor{black}{
Our study involves major  simplifications related to the relationship between diffusive and advective terms. We focus on scenarios for which a drop impacts and deforms primarily due to inertia. Consequently, we examine cases where the transport of surfactant is predominantly driven by advection rather than diffusion. This scenario corresponds to a Peclet number $Pe\gg 1$, allowing us to neglect certain diffusive terms and thereby limit the model complexity.
}

\textcolor{black}{
Despite several studies discussing the effects of surfactant diffusion along an interface \citep{manikantan2020surfactant}, understanding the associated mechanisms in dynamic situations remains limited. Determining a appropriate value for this diffusion coefficient for practical applications presents a significant challenge. A previous numerical study by \cite{antritter2019spreading} investigated the slow spreading of a surfactant-laden drop, which corresponds to a much lower Peclet number than we are concerned with; despite their much lower Peclet number regime, they omitted tangential diffusion along the interface from their computational model.  Taken together, both the difficulty in finding a physically relevant diffusion coefficient and this recent exclusion of diffusion in cases with a lower Peclet number support our decision to disregard the diffusive term in (\ref{IJMF2023:eqn:c_I}) for our numerical model} \textcolor{black}{by assuming that $D_\lambda$ = 0}.

\textcolor{black}{
Another diffusive term discussed in \cite{karapetsas2011surfactant} involves the diffusion of surfactant at the triple line from the liquid-air interface onto the substrate. For reasons related to the high $Pe$, we also disregard this effect. Instead, we consider only the sorption of surfactant at the contact line caused by the liquid flow, which forces surfactant at the liquid-air interface to come into contact with the substrate. This phenomenon is particularly significant when the falling drop first encounters the substrate, as a substantial portion of the drop interface meets the substrate. The forced contact is modelled by a vertical flux from the triple line to the substrate using}
\begin{equation}
\label{IJMF2023:eqn:surfaceSink}
j_{\textrm{IW}}= \max \left( - k \left(\boldsymbol{q} \cdot \boldsymbol{n}\textrm{w} \right) \delta\textrm{w} c_\textrm{I},0 \right),
\end{equation}
where $k$ is a coefficient of efficacy, which we set to 1, $\boldsymbol{n}_\textrm{w}$ is the normal unit vector of the substrate, and $\delta_\textrm{w}$ is similar to $\delta_\textrm{I}$ but evaluated for the substrate interface. The value of $\delta_\textrm{w}$ is non-zero only for cells adjacent to the substrate and is equal to $h/2$ for a regular hexahedral mesh. Equation (\ref{IJMF2023:eqn:surfaceSink}) acts as a sink for $c_\textrm{I}$ only when the velocity has a downward component.

\textcolor{black}{
Using the simplifications stated above, we can rewrite the governing equations for $c_{\textrm{I}}$ and $c_{\textrm{W}}$, namely (\ref{IJMF2023:eqn:c_I}) and (\ref{IJMF2023:eqn:c_W}), as
}
\begin{equation}
    \label{IJMF2023:eqn:c_I_mod}
    \frac{\partial c_{\textrm{I}}}{\partial t}
    + \grad \cdot \left( c_{\textrm{I}} \boldsymbol{q} \right)
    + \grad \cdot \left( c_{\textrm{I}} \left( 1 - 2 \mathcal{H} \right)\boldsymbol{q}_{\textrm{r}} \right)
    =
    j_{\textrm{IB}} + j_{\textrm{IW}},
\end{equation}
and
\begin{equation}
    \label{IJMF2023:eqn:c_W_mod}
    \frac{\partial c_{\textrm{W}}}{\partial t}
    =
    j_{\textrm{WI}}.
\end{equation}

\subsubsection{Additional corrections}

Equations (\ref{IJMF2023:eqn:c_B}) and (\ref{IJMF2023:eqn:c_I_mod}) both contain an artificial compression term that constrains $c_{\textrm{B}}$ to remain in the bulk and $c_{\textrm{I}}$ to remain at the interface.  However, for high concentrations of $c_{\textrm{B}}$ and $c_{\textrm{I}}$, we found these terms do not effectively confine the surfactants within their designated regions. To address this issue, we introduce an additional step to effectively restrict the movement of the surfactants.
A correction is applied on $c_B$ using
\begin{equation}
    \label{IJMF2023:eqn:c_BCorr}
    \frac{\partial c_{\textrm{B}}}{\partial \tau}
    =
    \grad^2 \left( c_{\textrm{B}} D_{\textrm{c}} \boldsymbol{T}_{\textrm{n}} \left( 1 - \mathcal{H} \right)   \right)
    - \grad \cdot \left( c_{\textrm{B}}  D_{\textrm{c}} \left( \frac{1 - \mathcal{H}}{\mathcal{H}} \right) \grad \mathcal{H} \right)
    ,
\end{equation}
where $\tau$ is an artificial time step with the same duration as $t$, $D_{\textrm{c}}= h \norm{\boldsymbol{q}}$ is an artificial diffusion coefficient that depends on the mesh size $h$ and the magnitude of the flux $\boldsymbol{q}$, and $\boldsymbol{T}_{\textrm{n}} =\boldsymbol{n} \otimes \boldsymbol{n} $ is a tensor. This correction term is only active at the interface where the $c_B$ values go from the local peripheral bulk concentration to 0; it is designed to maintain $c_{\textrm{B}} \mathrm{/} \mathcal{H}$ constant in the normal direction of the interface.
The correction applied to $c_{\textrm{I}}$ is
\begin{equation}
    \label{IJMF2023:eqn:c_ICorr}
    \frac{\partial c_{\textrm{I}}}{\partial \tau}
    =
    \grad^2 \left( c_{\textrm{I}} D_{\textrm{c}} \boldsymbol{T}_{\textrm{n}} \left( 1 - \delta_{\textrm{I}} \varepsilon \right)   \right)
    - \grad \cdot \left( c_{\textrm{I}}  D_{\textrm{c}} \left( \frac{1 - \delta_{\textrm{I}} \varepsilon}{\delta_{\textrm{I}}} \right) \grad \mathcal{H} \right).
\end{equation}
In this case, the correction is designed to maintain $c_{\textrm{I}} \mathrm{/} \delta_{\textrm{I}}$ or $\Gamma$ constant in the normal direction of the interface and helps obtain a constant surface-specific surfactant concentration in the normal direction of the interface while preserving the gradient along the interface.

\subsubsection{Source-sink terms at the liquid-gas interface} \label{IJMF2023:sec:newModLH}

The source-sink terms $j_{\textrm{IB}}$ and $j_{\textrm{BI}} $ at the liquid-gas interface depend on the imbalance between the concentration of surfactant at the interface $c_\textrm{I}$ and the concentration of surfactant in the bulk next to the interface $c_{\textrm{B}\shortrightarrow\textrm{I}}$, also called subsurface concentration, which is the only concentration in the bulk that interacts with the interface.

The kinetics of the sorption model also depends on $\Gamma \textrm{/} \Gamma_\textrm{m}$, where $\Gamma$ represents the surface-specific concentration of surfactants at the interface, and $\Gamma_\textrm{m}$ denotes its maximum value. The maximum value of $\Gamma$ is a constant parameter that is dependent on the properties of the solvent and surfactants. The surface-specific concentration at the interface $\Gamma$ is calculated using
\begin{equation}
    \label{IJMF2023:eqn:Gamma}
    \Gamma = \frac{c_{\textrm{I}}}{\delta_{\textrm{I}}},
\end{equation}
where $\delta_{\textrm{I}}$ represents the quantity of interface area per unit volume, as defined in Section \ref{IJMF2023:sec:LS}.

To account for the energy barrier that occurs for a high concentration of surfactants at the interface, a modified Langmuir-Hinshelwood kinetic model \citep{chang1992modified} is used to obtain the source/sink term $j_{\textrm{IB}}$:
\begin{equation}
    \label{IJMF2023:eqn:modLH}
    j_{\textrm{IB}}
    =
    \exp\left(-B M\right)
    k_{\textrm{ad}} c_{\textrm{B}\shortrightarrow\textrm{I}}
    \left( \delta_{\textrm{I}} - \frac{c_{\textrm{I}}}{\Gamma_{\textrm{m}}}\right)
    -
    \exp\left(-B M\right)
    k_{\textrm{de}} c_{\textrm{I}},
\end{equation}
where $k_{\textrm{ad}}$ and $k_{\textrm{de}}$ denote the adsorption and desorption rate coefficients, respectively, while $B$ is an empirical parameter and $M$ is a dimensionless quantity that is typically calculated from $\Gamma \mathrm{/} \Gamma_\textrm{m}$.
The exponential terms in (\ref{IJMF2023:eqn:modLH}) are responsible for slowing down the flux when the surfactant concentration at the liquid-air interface is high. This is due to the creation of an activation energy barrier caused by the surfactant molecules that are already adsorbed at the interface.
It is worth noting that the coefficients $k_{\textrm{ad}}$, $k_{\textrm{de}}$, and $B$ in (\ref{IJMF2023:eqn:modLH}) are influenced by the concentration of surfactants in the bulk solution. However, we make the assumption that the changes in $c_{\textrm{B}}$ are negligible, allowing us to treat $k_{\textrm{ad}}$, $k_{\textrm{de}}$, and $B$ as constants.

The area-specific concentration at equilibrium $\Gamma_{\textrm{eq}}$ is determined using (\ref{IJMF2023:eqn:Gamma}) and (\ref{IJMF2023:eqn:modLH}) with $j_{\textrm{IB}}= 0$ via
\begin{equation}
    \label{IJMF2023:eqn:Gamma_eq}
    \Gamma_{\textrm{eq}}=\frac{k_{\textrm{ad}} c_{\textrm{B}\shortrightarrow\textrm{I}}}{k_{\textrm{ad}} c_{\textrm{B}\shortrightarrow\textrm{I}} + k_{\textrm{de}} \Gamma_{\textrm{m}}} \Gamma_{\textrm{m}}.
\end{equation}
The Langmuir isotherm $K_\textrm{L}$, which is a constant that depends on the surfactant, relates the adsorption and desorption rate coefficients and the maximum possible adsorbed amount: 
\begin{equation}
    \label{IJMF2023:eqn:LangmuirIso}
    K_\textrm{L} = \frac{k_{\textrm{ad}}}{k_{\textrm{de}} \Gamma_{\textrm{m}}}.
\end{equation}
Therefore, the area-specific concentration at equilibrium can be re-written as
\begin{equation}
    \label{IJMF2023:eqn:Gamma_eqK}
\Gamma_{\textrm{eq}}=\frac{c_{\textrm{B}\shortrightarrow\textrm{I}}}{c_{\textrm{B}\shortrightarrow\textrm{I}} + \frac{1}{K_\textrm{L}}} \Gamma_{\textrm{m}}.
\end{equation}
Equation (\ref{IJMF2023:eqn:Gamma_eqK}) shows that $\Gamma_{\textrm{eq}}$ depends on two surfactant properties, which are the constants $K_\textrm{L}$ and $\Gamma_\textrm{m}$, and the concentration $c_{\textrm{B}\shortrightarrow\textrm{I}}$.

The maximum equilibrium surface-specific concentration $\Gamma_{\textrm{eq}}^\textrm{CMC}$ is obtained for solutions with a concentration of surfactants equal to or above the CMC. To satisfy this constraint, two adjustments need to be made. Firstly, the value of the subsurface concentration cannot exceed the CMC, \ie, $c_{\textrm{B}\shortrightarrow\textrm{I}} \in (0,\textrm{CMC})$. Secondly, (\ref{IJMF2023:eqn:modLH}) includes a barrier term, which needs to be decreased when the surface concentration exceeds $\Gamma_{\textrm{eq}}^\textrm{CMC}$, allowing surfactants to desorb. To this end, a modified factor $M$ is introduced, as given by
\begin{equation}
\label{IJMF2023:eqn:barrierExpMod}
M=
    \begin{cases}
    \frac{\Gamma^\textrm{s}}{\Gamma_\textrm{m}}
    &\textrm{if $\Gamma^\textrm{s}< \Gamma_{\textrm{eq}}^\textrm{CMC}$}\\
    \max \left(
        \frac{\Gamma_{\textrm{eq}}^\textrm{CMC}}{\Gamma_\textrm{m}} - C \frac{\Gamma^\textrm{s}-\Gamma_{\textrm{eq}}^\textrm{CMC}}{\Gamma_\textrm{m}}
    ,0\right)
    &\textrm{if $\Gamma^\textrm{s} > \Gamma_{\textrm{eq}}^\textrm{CMC}$}
    \end{cases}
    ,
\end{equation}
where $\Gamma^\textrm{s}$ is a smoothed value of $\Gamma$ and $C=5$ is an empirical parameter that facilitates desorption when $\Gamma^\textrm{s} > \Gamma_{\textrm{eq}}^\textrm{CMC}$. To enhance numerical stability, the smoothed variable $\Gamma^\textrm{s}$ is utilised instead of the conventional variable $\Gamma$. The smoothed value of $\Gamma^\textrm{s}$ is obtained using equation (\ref{IJMF2023:eqn:Gamma}) with the smoothed volume-specific concentration $c_\textrm{I}^\textrm{s}$ and the smoothed delta function $\delta_\textrm{I}^\textrm{s}$ are employed. The smoothing operation is achieved by a ``cell-to-point-to-cell interpolation'' followed by diffusion in the normal direction of the interface. \textcolor{black}{The adjusted barrier term $M$ in equation (\ref{IJMF2023:eqn:barrierExpMod}) plays a vital role in preventing the interface concentration from approaching values near $\Gamma_\textrm{m}$, which could result in a nonphysical negative surface tension.}

Due to the spatial displacement between the subsurface and the interface, the source-sink terms were calculated explicitly in \cite{antritter2019spreading}, which does not cause numerical issues in their study because they consider liquids with a low concentration of surfactants. In our study where the concentration of surfactants can be high, using explicit source-sink terms can cause stability issues when solving the equations (\ref{IJMF2023:eqn:c_B}) and (\ref{IJMF2023:eqn:c_I}).
To address this issue, we develop a method where $j_{\textrm{BI}}$ in (\ref{IJMF2023:eqn:c_B}) and $j_{\textrm{IB}}$ in (\ref{IJMF2023:eqn:c_I}) are split into
an explicit and an implicit part: $j_{\textrm{BI}} = S_\textrm{BI}^\textrm{Im} c_\textrm{B} + S_\textrm{BI}^\textrm{Ex}$ and $j_{\textrm{IB}} = S_\textrm{IB}^\textrm{Ex} + S_\textrm{IB}^\textrm{Im} c_\textrm{I}$ where $S_\textrm{IB}^\textrm{Ex}$ and $S_\textrm{BI}^\textrm{Ex}$ are the explicit source-sink sub-terms and $S_\textrm{IB}^\textrm{Im}$ and $S_\textrm{BI}^\textrm{Im}$ the implicit source-sink sub-terms. The implicit terms enhance the numerical stability, but they require establishing a direct link between the subsurface and the interface to exchange information between them.
To accomplish this, we locate the two closest cells in the subsurface for each cell of the interface, or the nearest group of cells in the subsurface if they are equidistant from the interface cell. These links allow the transmission of implicit information from the interface to the subsurface and vice versa. Multiple links are formed between cells from the interface and subsurface and the weight of each link depends on the number of connections. Typically, neighbouring cells share one or more common connections, which results in local overlap, enabling us to smooth the  sink-source terms and enhance stability. Due to the links, $S_\textrm{BI}^\textrm{Im}$ and $S_\textrm{BI}^\textrm{Ex}$ are the non zeros only in the subsurface, while $S_\textrm{IB}^\textrm{Ex}$ and $S_\textrm{IB}^\textrm{Im}$ are non-zeros at liquid-gas interface. Using the A-CVOFLS, we determine that the interface is composed of cells with $\lvert \psi \rvert <\varepsilon$.

In order to take into account the concentration limit of $c_{\textrm{B}\shortrightarrow\textrm{I}}$ while using explicit and implicit sink-source terms, we rewrite $j_{\textrm{BI}}$ as
\begin{equation}
\label{IJMF2023:eqn:sink}
j_{\textrm{BI}}
=
\underbrace{
-\exp\left(-B M\right)
k_{\textrm{ad}}
\left(-\delta_\textrm{I}
+\frac{c_{\textrm{I}}}{\Gamma_{\textrm{m}}}\right)
c_{\textrm{B}}
}_{\mathclap{\text{implicit}}}
+
\underbrace{
\vphantom{\left(\left(\frac{c_{\textrm{I}}}{\Gamma_{\textrm{m}}}\right)\right)_L}
\exp\left(-B M\right)
\left(
k_{\textrm{de}} c_{\textrm{I}}
-
k_{\textrm{ad}}
\left(-\delta_\textrm{I}
+\frac{c_{\textrm{I}}}{\Gamma_{\textrm{m}}}\right)
\max(c_\textrm{B} - \textrm{CMC}, 0)
\right)
}_{\mathclap{\text{explicit}}}
,
\end{equation}
and $j_{\textrm{IB}}$ as
\begin{equation}
\label{IJMF2023:eqn:source}
j_{\textrm{IB}}
=
\underbrace{
\vphantom{\left(\frac{k_{\textrm{ad}}}{\Gamma_{\textrm{m}}}\right)_L}
\exp\left(-B M\right)
k_{\textrm{ad}} \delta_\textrm{I}
\min(c_{\textrm{B}},\textrm{CMC})
}_{\mathclap{\text{explicit}}}
+
\underbrace{
\exp\left(-B M\right)
\left(
-\frac{k_{\textrm{ad}}}{\Gamma_{\textrm{m}}}
c_{\textrm{B}}
- k_{\textrm{de}}
\right)
c_{\textrm{I}}
}_{\mathclap{\text{implicit}}}
,
\end{equation}
where the max and min functions are used to prevent the micelles from affecting the sorption dynamics.

\subsubsection{Surface tension}

The surface tension of the liquid $\sigma$ deviates from the solvent surface tension $\sigma_{\textrm{0}}$ depending on the amount of adsorbed surfactant at the liquid-gas interface.
\textcolor{black}{There exist various possible equations of state to describe this dependence, each with different degrees of complexity.  Notable among these are the Langmuir and Frumkin equations of state. For the purposes of this paper, we have chosen to use the Langmuir equation due to its relative simplicity and its suitability for the specific surfactant used in our simulations, which will be presented later.  The relevant formula is}
\begin{equation}
    \label{IJMF2023:eqn:sigma}
    \sigma=\sigma_{\textrm{0}}
    + n R T \Gamma_{\textrm{m}}
    \ln{\left( 1-\frac{\Gamma^{\textrm{s}}}{\Gamma_{\textrm{m}}} \right)},
\end{equation}
where $n$ is a constant that depends on the ionic nature of the surfactant, $R$ is the universal gas constant and $T$ is the temperature in Kelvin.
While $\Gamma^{\textrm{s}}$ represents a smoothed variant of $\Gamma$, it is important to note that the smoothing procedure operates perpendicular to the interface. This perpendicular smoothing process is crucial as it maintains the gradient of surfactant concentration along the interface, thereby preserving the gradient of surface tension in Equation~(\ref{IJMF2023:eqn:STF}).
\textcolor{black}{As a reminder, $\Gamma^{\textrm{s}}$ will remain comparatively small compared to ${\Gamma_{\textrm{m}}}$, due to the modified barrier term introduced in (\ref{IJMF2023:eqn:barrierExpMod}), which prevents the surface tension from reaching nonphysical negative values.}


\color{black}

\subsection{Summary of the CFD model}

At each time step, the governing equations for the two-phase flow are solved; these equations include the mass conservation
 (incompressibility) (\ref{IJMF2023:eqn:incompressible}), the momentum conservation  (\ref{IJMF2023:eqn:NavierStokes}) and, the volume fraction advection (\ref{IJMF2023:eqn:alphaAdvMod}). An adaptive compression term (\ref{IJMF2023:eqn:adaptComp}) and a modified A-CLSVOF method are used in this process.

Following the two-phase flow calculation, the governing equations for the surfactants are solved. These equations are specific to the surfactants in the bulk (\ref{IJMF2023:eqn:c_B}), at the liquid-air interface (\ref{IJMF2023:eqn:c_I_mod}), and on the substrate (\ref{IJMF2023:eqn:c_W_mod}). These equations are coupled using source-sink terms (\ref{IJMF2023:eqn:surfaceSink}), (\ref{IJMF2023:eqn:sink})-(\ref{IJMF2023:eqn:source}). After solving the surfactant equations, the solutions for the surfactant concentration in the bulk and at the interface are corrected using equations (\ref{IJMF2023:eqn:c_BCorr}) and (\ref{IJMF2023:eqn:c_ICorr}), respectively. Finally, the surface tension is updated using the newly corrected surfactant concentration at the liquid-air interface.

\color{black}

\section{Results and discussions} \label{IJMF2023:sec:results}

In this section, we employ the model to simulate the impact of drops that contain surfactants. The simulations serve two primary purposes: to showcase the model's potential in analysing the impact dynamics of surfactant-laden drops and also to highlight its inherent limitations.

To investigate the impact dynamics influenced by surfactants, we conduct two key analyses. Firstly, we compare the impact of a surfactant-laden reference liquid and its pure solvent counterpart, at different impact velocities (Section~\ref{IJMF2023:sec:SaaDist}). Secondly, we conduct a sensitivity analysis by varying parameters of the reference liquid (Section~\ref{IJMF2023:sec:sensiAna}).

In this study, all the simulated drops are initially spherical with a diameter of 1 \si{\milli\meter}, made of water, and generated at 0.1 \si{\milli\meter} above the substrate. Initially, the surfactant concentration at the interface is in equilibrium with the bulk concentration. The drops impact the substrate with velocities ranging from 0.5 to 2 \si{\meter\per\second}. During the drop spreading, the contact angle follows the Kistler model (Section~\ref{IJMF2023:sec:Kistler}), with an input of an advancing contact angle of 90\si{\degree}.

The simulations are performed using an axisymmetric 1.8 $\times$ 1.2 \si{\milli\meter} wedge domain with an opening angle of 5\si{\degree}. The domain is discretised into a regular grid of 120 $\times$ 80 cells with a width of 0.015 \si{\milli\meter}. To accurately capture the details of the liquid-gas interface, the dynamic mesh is adaptive and is refined three times in the vicinity of the interface (see Section~\ref{IJMF2023:sec:adaptMesh}).

\subsection{Surfactant distribution} \label{IJMF2023:sec:SaaDist}

The surfactant model outlined in Section \ref{IJMF2023:sec:compWorkSaa} involves multiple variables. To use values that are physically meaningful in practical applications, we opted to employ the parameters associated with a widely-used surfactant, namely sodium dodecyl sulfate (SDS).
The surfactant has a diffusion coefficient $D_\textrm{B}=$ \num{7.84e-11} \si{\square\meter\per\second}, Langmuir's isotherm $K_\textrm{L}=$ \num{1.1e-1} \si{\cubic\meter\per\mol}, critical micelle concentration $\textrm{CMC}=$ \num{8.1e-3} \si{\mol\per\cubic\meter}, and maximum surface-specific concentration $\Gamma_\textrm{m}=$ \num{4.7e-6} \si{\mol\per\square\meter}, a temperature $T=$ 298\si{\kelvin}, and a constant $n=$ 2.
The reference solution has an initial concentration of surfactant $c_B^0$ = \num{8.1e-3} \si{\mol\per\cubic\meter}, which is equivalent to the critical micelle concentration. This value implies that the dimensionless barrier parameter $B$ is equal to 32 and the adsorption rate coefficient $k_\textrm{ad}$ is 3 \si{\meter\per\second} \citep{chang1992modified}.

\begin{figure}
\centering\includegraphics[width=0.5\linewidth]{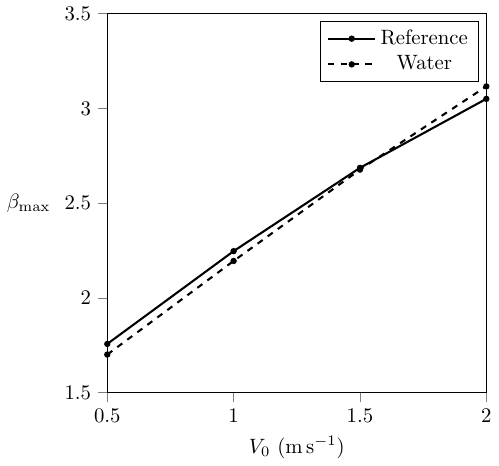}
\caption{Relative maximum spreading diameters ($\beta_\textrm{max}$) for reference solution and pure water at various impact velocities ($V_0$). Faster $V_0$ results in larger $\beta_\textrm{max}$ for both solutions. The difference in $\beta_\textrm{max}$ between the two solutions is caused by their respective interfacial forces.}
\label{IJMF2023:fig:2}
\end{figure}

Figure~\ref{IJMF2023:fig:2} compares the relative maximum spreading diameters between the reference solution ($\beta_\textrm{max}^\textrm{Ref}$) and pure water ($\beta_\textrm{max}^\textrm{Water}$) under different impact velocities. At low impact velocities ($V_0 \approx 0.5-1$ \si{\meter\per\second}), both $\beta_\textrm{max}$ values show a linear increase with respect to $V_0$ and maintain a parallel relationship, with $\beta_\textrm{max}^\textrm{Ref}>\beta_\textrm{max}^\textrm{Water}$. As the impact velocity increases, the maximum spreading diameters of both solutions continue to increase monotonically, but the rate of increase for $\beta_\textrm{max}^\textrm{Ref}$ becomes lower than that of $\beta_\textrm{max}^\textrm{Water}$. Consequently, for higher impact velocities ($V_0 \approx 1.5-2$ \si{\meter\per\second}), $\beta_\textrm{max}^\textrm{Ref}$ falls below $\beta_\textrm{max}^\textrm{Water}$. These observations indicate that the impact behaviour of the surfactant-laden drops deviates from the drops made from the solvent as the impact velocity increases. This deviation in their $\beta_\textrm{max}$ behaviour is attributed to the presence of surfactants, which alter the interfacial forces depending on the impact velocity.

\begin{figure}
  \centering
  \begin{subfigure}[t]{0.6\linewidth}
    \centering
    \includegraphics[width=1\linewidth]{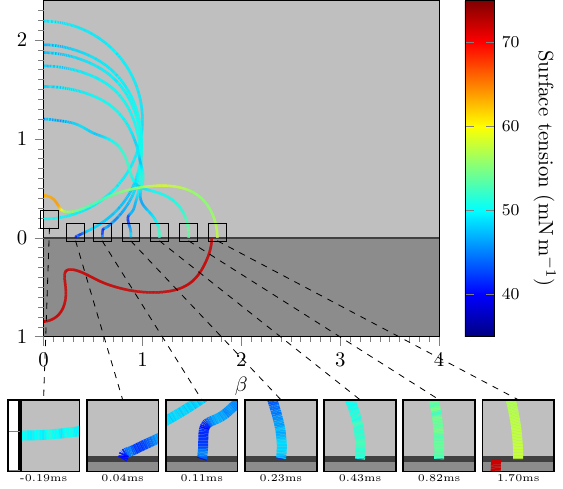}
    \caption{Impact velocity of 0.5 \si{\meter\per\second}.}
  \end{subfigure}
  \vspace{0.5cm}
  \begin{subfigure}[b]{0.6\linewidth}
    \centering
    \includegraphics[width=1\linewidth]{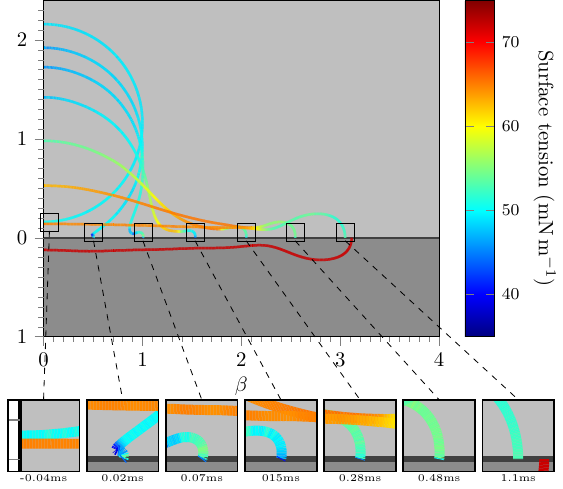}
    \caption{Impact velocity of 2.0 \si{\meter\per\second}.}
  \end{subfigure}
  \caption{Simulation of a 1 \si{\milli\meter} aqueous drop containing a concentration of \num{8.1e-3} \si{\mol\per\cubic\meter} (=CMC) impacting on a flat surface up to its maximum diameter with a velocity of (a) 0.5 \si{\meter\per\second} and (b) 2 \si{\meter\per\second}. The maximum spreading of the surfactant-laden drop is compared to a pure water drop (solvent), which is displayed under the horizontal by changing the sign of the interface y values. The surface tension is displayed using a colour map.}
  \label{IJMF2023:fig:3}
\end{figure}

The surface tension of a liquid-gas interface is influenced by the concentration and distribution of surfactants, as indicated in (\ref{IJMF2023:eqn:sigma}), which in turn affects the interfacial forces, according to (\ref{IJMF2023:eqn:STF}).
This phenomenon is demonstrated in Figure \ref{IJMF2023:fig:3}, which provides a comprehensive depiction of the dynamic changes in surface tension observed in a simulated surfactant-laden drop during impact at two distinct velocities: (a) a relatively low impact velocity of $V_0=0.5$ \si{\meter\per\second}, and (b) a higher impact velocity of $V_0=2.0$ \si{\meter\per\second}.
To capture the evolution of the drop's shape for each impact velocity, seven distinct profiles, characterised by $\alpha=0.5$, are depicted at different time intervals.
These time points span from just prior to the drop's contact with the flat substrate to the moment when the drop reaches its maximum spreading diameter. Note the time is set to zero when the drop initiates its spreading.
The colour-coding of these profiles corresponds to the varying surface tension values, as depicted by the colour bar.
Furthermore, each profile includes an inset that offers a magnified view of the region near the contact line.
Lastly, a single inverted profile is included beneath the horizontal axis, representing the profile of a drop composed solely of the solvent at its maximum spreading diameter, serving as a useful benchmark for comparison.
To enhance clarity, we have omitted the representation of the small air bubble that forms at the impact point.

At a low impact velocity of 0.5 \si{\meter\per\second}, the surfactant-laden drop spreads to achieve $\beta_\textrm{max} < 2$ within 1.7 \si{\milli\second} (see Fig.~\ref{IJMF2023:fig:3}(a)). During impact, the surface tension evolves from approximately 49 \si{\milli\newton\per\meter} ($\approx \Gamma_\textrm{eq}$) before impact to around 55 \si{\milli\newton\meter} for the majority of the liquid-gas interface when the drop reaches its maximum diameter. As the drop spreads, the distribution of surfactants causes portions of the interface to experience a decrease in surface tension due to the accumulation of surfactants, while other portions experience an increase due to the expansion of the interfacial area. Despite these localized effects, the overall result is an expansion of the interfacial area, leading to a higher surface tension globally. However, the observed increase of surface tension remains significantly lower than the solvent's surface tension.  As a result, $\beta_\textrm{max}^\textrm{Ref}$ is greater than $\beta_\textrm{max}^\textrm{Water}$.

At a higher velocity of 2 \si{\meter\per\second}, the surfactant-laden drop deforms and reaches $\beta_\textrm{max} > 2$ in about 1.1 \si{\milli\second} (Fig.~\ref{IJMF2023:fig:3}(b)). This deformation is larger and faster compared to that observed at $V_0 = 0.5$, resulting in an increase of surface tension from around 49 \si{\milli\newton\per\meter} to 65 \si{\milli\newton\per\meter} for a significant part of the liquid-gas interface. This increase in surface tension is caused by the expansion of the interface, which starts from the outer part of the drop (excluding the rim) and propagates toward the central part. However, the surface tension at the leading edge remains low compared to the rest of the interface because of the accumulation of surfactants, which was assumed from experimental results by \cite{zhang1997dynamic}. The lower surface tension at the rim results in a gradient of surface tension that generates Marangoni forces that oppose the drop spreading. The combination of this effect with the global increase in surface tension leads to a lower value of $\beta_\textrm{max}^\textrm{Ref}$ compared to $\beta_\textrm{max}^\textrm{Water}$.

\subsection{Surfactant sorption} \label{IJMF2023:sec:sensiAna}

The deformation of the drop during impact alters the concentration of surfactants at the liquid-gas interface, causing it to deviate from its equilibrium state. The extent to which this phenomenon can be mitigated by the rate of sorption depends on the properties of the surfactant. Faster sorption of surfactants at the liquid-gas interface limits the increase in surface tension and the rise of Marangoni forces that are generated during the spreading of the surfactant-laden drops. To understand the influence of surfactant on the maximum diameter, we conducted a sensitivity analysis by varying the barrier coefficient ($B$), the adsorption coefficient ($k_\textrm{ad}$), and the initial concentration of surfactant in the drop ($c_B^0$). The properties of all simulated cases are detailed in Table~\ref{IJMF2023:tab:simuPara}. The modified Langmuir-Hinshelwood kinetic model (Eq.~\ref{IJMF2023:eqn:modLH}), as outlined in Section~\ref{IJMF2023:sec:newModLH}, specifies that $B$ and $k_\textrm{ad}$ depend on $c_B^0$. As a result, we conducted separate assessments of the effects of $B$ and $k_\textrm{ad}$ before investigating these parameters across varying $c_B^0$.
\begin{table}
\centering
\caption{Parameters for the simulated cases.}
\label{IJMF2023:tab:simuPara}
\begin{threeparttable}
\begin{tabular}{llll}
\hline
Name&      $c_B^0$&    $B$&     $k_\textrm{ad}$\\
    &\si{\mol\per\cubic\meter}&    &     \si{\meter\per\second}\\
\hline
Reference\tnote{*}&        \num{8.1e-3}&        32&     3\\
$B_{22}$&   \num{8.1e-3}&        22&     \num{3}\\
$B_{11}$&   \num{8.1e-3}&        11&     \num{3}\\
$K_{30}$&   \num{8.1e-3}&        32&     \num{30}\\
$K_{300}$&  \num{8.1e-3}&        32&     \num{300}\\
$C_{5.9}$\tnote{*}&  \num{5.9e-3}&        22&     \num{2.8e-2}\\
$C_{3.3}$\tnote{*}&  \num{3.3e-3}&        22&     \num{9e-4}\\
$C_{1.7}$\tnote{*}&  \num{1.7e-3}&        52&     \num{5.5e-4}\\
\hline
\end{tabular}
\begin{tablenotes}\footnotesize
\item[*] Values taken from \citep{chang1992modified}
\end{tablenotes}
\end{threeparttable}
\end{table}

\begin{figure}
\centering\includegraphics[width=1\linewidth]{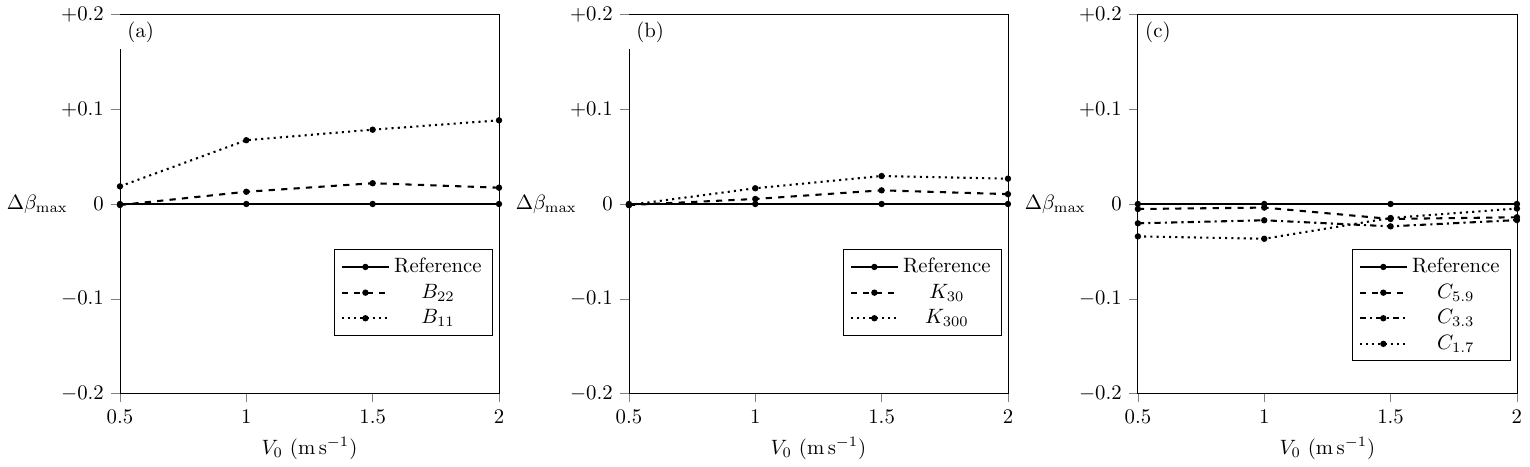}
\caption{Difference of maximum relative spreading diameter between the alternative solutions and the reference solution for different impact velocities.}
\label{IJMF2023:fig:4}
\end{figure}

Figure~\ref{IJMF2023:fig:4} depicts the change in the maximum relative spreading diameter, $\Delta\beta_\textrm{max}$, between the reference solution and the alternative solutions for various impact velocities.
To investigate the impact of changes in the barrier coefficient $B$, we compared liquids $B_{11}$ and $B_{22}$ with the reference solution ($B=32$, Tab.~\ref{IJMF2023:tab:simuPara}), as shown in Figure~\ref{IJMF2023:fig:4}(a). A decrease in $B$ allows for faster adsorption of surfactants at the liquid-gas interface for the same $\Gamma$ when $\Gamma < \Gamma_\textrm{eq}$, which is the case for most of the interface during drop spreading. Lowering $B$ limits the increase in surface tension by promoting the adsorption of surfactants and preventing the rise of strong Marangoni forces, particularly at high-impact velocities. Overall, a lower $B$ results in a higher $\beta_\textrm{max}$ as demonstrated in Figure~\ref{IJMF2023:fig:4}a where the largest value of $\beta_\textrm{max}$ is obtained by for $B_{11}$, followed by $B_{22}$.

To examine the effect of changes in the adsorption coefficient $k_\textrm{ad}$, we compared liquids $K_{30}$ and $K_{300}$ with the reference solution ($k_\textrm{ad}=3$, Tab.~\ref{IJMF2023:tab:simuPara}), as shown in Figure~\ref{IJMF2023:fig:4}b. An increase in $k_\textrm{ad}$ leads an increase of $k_\textrm{de}$ (see Eq.~\ref{IJMF2023:eqn:LangmuirIso}), which means that the equilibrium surface tension remains the same. As with decreasing $B$, increasing $k_\textrm{ad}$ limits surface tension and Marangoni stress by promoting the adsorption of surfactants. However, unlike $B$, which appears in the exponential term in the sink-source term in (\ref{IJMF2023:eqn:modLH}), $k_\textrm{ad}$ linearly affects $j_\textrm{IB}$, resulting in $\beta_\textrm{max}$ being less sensitive to $k_\textrm{ad}$ than $B$.

\begin{figure}
\centering\includegraphics[width=1\linewidth]{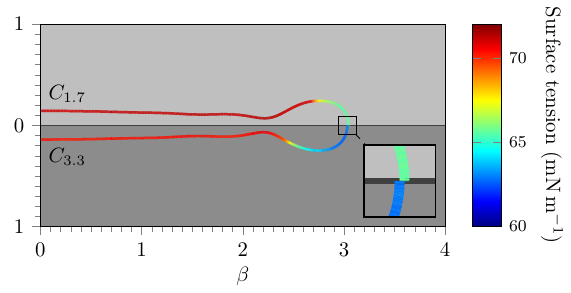}
\caption{Simulation of two 1 \si{\milli\meter} aqueous drops with different concentrations of surfactants (\num{1.7e-3} and \num{3.3e-3} \si{\mol\per\cubic\meter}) impacting on a flat surface with a velocity of 2 \si{\meter\per\second}. Their maximum spreading is compared, with the solution containing \num{1.7e-3} \si{\mol\per\cubic\meter} that is displayed above the horizontal while the solution containing \num{3.3e-3} \si{\mol\per\cubic\meter} that is displayed under the horizontal. The surface tension is displayed using a colour map.}
\label{IJMF2023:fig:5}
\end{figure}

To investigate the effect of changes in the initial solution concentration $c_\textrm{B}^0$, we compared liquids $C_{1.7}$, $C_{3.3}$, and $C_{5.9}$ with the reference solution ($c_\textrm{B}^0=$ \num{8.1e-3} \si{\mol\per\cubic\meter}, Tab.\ref{IJMF2023:tab:simuPara}), as shown in Figure~\ref{IJMF2023:fig:4}c. The equilibrium surface tension is closer to the solvent for liquids with lower initial surfactant concentration. Therefore, even before impact, the surface tension of $C_{1.7}$, $C_{3.3}$, and $C_{5.9}$ is higher than that of the reference solution. At low-impact velocities, the liquids with lower $c_\textrm{B}^0$ have a lower $\beta_\textrm{max}$, consistent with the fact that drop impact dynamics are primarily related to initial surface tension. At high-impact velocities, the relationship between the initial concentration and $\beta_\textrm{max}$ becomes more intricate. Figure~\ref{IJMF2023:fig:5} illustrates the surface tension of a drop containing \num{1.7e-3} \si{\mol\per\cubic\meter} and \num{3.3e-3} \si{\mol\per\cubic\meter} at their maximum diameter upon impact at a velocity of 2 \si{\meter\per\second}. For approximately the same $\beta_\textrm{max}$, $C_{1.7}$ exhibits a higher overall surface tension compared to $C_{3.3}$. However, due to the steeper surface tension gradient of $C_{3.3}$ resulting in stronger Marangoni forces, its $\beta_\textrm{max}$ is slightly lower than that of $C_{1.7}$, despite having a lower overall surface tension. This example demonstrates that at higher impact velocities, the influence of $c_\textrm{B}^0$ on interfacial forces has an opposite effect on $\beta_\textrm{max}$: higher surfactant concentrations lead to lower surface tension forces, which increase $\beta_\textrm{max}$, while higher surfactant concentrations result in stronger Marangoni forces caused by steeper surface tension gradient, leading to a decrease in $\beta_\textrm{max}$. The combined effect of these two opposing phenomena determines the value of $\beta_\textrm{max}$.

\section{Conclusions} \label{IJMF2023:sec:ccl}

The main objective of this study is to present a comprehensive CFD model that simulates the impact of surfactant-laden drops on a flat surface.  As far as we are aware, ours is the only computational framework that uses the VOF method to meet this objective.
Unlike recent computational studies of drop spreading/receding \citep{badra2018numerical,antritter2019spreading,wang2022unconditionally} and impacting \citep{ganesan2015simulations,wang2022energy} that include the effect of surfactant, we focus specifically on cases in which the drop contains a high concentration of surfactant at its liquid-gas interface and in the bulk (below the critical micelle concentration).

Our numerical approach employs the volume-of-fluid method within the OpenFOAM framework, augmented by a level-set method that allows for a more accurate interface description. The most significant computational challenges arise from the inclusion of surfactant in the liquid, which ultimately leads to Marangoni forces along the interface that can resist the spreading. In particular, we have had to deal with complexity that arises from intricate interactions between surfactants within the droplet volume, at the liquid-gas interface, and at the liquid-solid interface while the liquid is spreading rapidly along the substrate. While our study exclusively focuses on drop impaction scenarios, we believe our CFD model could form the basis of numerical schemes for simulating other inertial-driven interfacial flows with surfactants and substrates.

By simulating various scenarios involving surfactant-laden drops with different properties, we have gained a deeper understanding of the distribution of surfactants at the liquid-air interface during drop spreading and their influence on interfacial forces, which in turn alter the maximum spreading and impact dynamics.
The higher the initial concentration, the greater the presence of surfactant at the liquid-gas interface, resulting in a decreased surface tension of the drop prior to impact. Upon impact, the overall drop surface tension rises and approaches the solvent's surface tension, as the surfactant becomes diluted due to the increase in interfacial area. This phenomenon is particularly noticeable at higher impact velocities. By increasing the adsorption rate, either through an increase in the adsorption coefficient or a decrease in the barrier coefficient, the rise in surface tension can be mitigated. The increase in surface tension is observed across most of the liquid-gas interface, except at the rim where surfactants accumulate through advection. The disparity in surface tension between the rim and the central part of the drop drives Marangoni forces that oppose drop spreading. To summarise, at lower impact velocities, the maximum spreading depends almost exclusively on the initial surface tension, whereas at higher impact velocities, maximum spreading depends on the overall increase in surface tension and the strength of Marangoni forces. Both phenomena reduce spreading and are weakened by a faster rate of sorption.
A broader observation is that the dominant forces that affect drop impaction are inertia and surface tension, and it turns out that the effects of Marangoni forces are subtle and rather difficult to isolate even within a computational framework.


Our computational model has some limitations that could be addressed to improve its accuracy and applicability. \textcolor{black}{Firstly, our numerical model is tailored  for instances with a high Peclet number, in which some diffusive terms are neglected}. Secondly, one of the model's constraints relates to the initial concentration of surfactant in the bulk, which is limited to the critical micelle concentration. Additionally, the model does not account for some of the surfactant transfers between the drop and the substrate.  Lastly, the Kistler model for the dynamic contact angle is parametric and not specifically designed to deal with drops containing surfactants. Overcoming these limitations would allow the model to better simulate drops containing a high concentration of micelles and analyse their impact dynamics while they spread and recede. Expanding the model in this way would be relevant in many industrial processes, such as agrichemical spraying.
Finally, an important future direction could involve extending the simulations to three dimensions, thereby enabling the study of surfactant-laden drops that undergo shattering events and interact with complex substrates, either fabricated \citep{broom2023water, khojasteh2016droplet} or natural \citep{dorr2015impaction,huet2020image}. Such three-dimensional simulations with sufficient accuracy would of course require substantial computational resources.

\section*{Acknowledgments}

This research was co-funded by the Queensland University of Technology (QUT) and Syngenta. The authors thank for their support the Australian Research Council Linkage Project LP160100707, in which Syngenta and Nufarm are the industry partners.
Computational resources and services used in this work were provided by HPC and Research Support Group, Queensland University of Technology, Brisbane, Australia.
The authors express their gratitude to Thomas Antritter,  Ian Turner, Alison Forster and Jerzy Zabkiewicz for insightful and productive  and discussions.


\section*{Appendix}

\subsection{Time step}

The number and complexity of the equations described in this study make the simulation computationally demanding. Using an adaptive time step allows the simulation to dynamically adjust the time step size based on the flow and fluxes conditions, ensuring that the simulation is accurate while minimising computational resources.
The time step, $\Delta t$, is adaptive and is automatically calculated respecting the strictest of the following conditions:
\begin{enumerate*}[label=(\roman*)]
    \item the Courant–Friedrichs–Lewy condition
    \item the Dehpande condition and
    \item the sorption condition.
\end{enumerate*}

The Courant–Friedrichs–Lewy (CFL) condition is based on the size of the computational grid and the velocity of the fluid being simulated.:
\begin{equation*}
    \label{IJMF2023:eqn:dT4Speed}
    \Delta t \leq C \max \left( \frac{h}{\norm{\boldsymbol{q}}} \right),
\end{equation*}
where $C$ is the Courant number. $C$ is usually set to 1 or lower to reduce the risk of instability.

Spurious currents play a secondary role in the context of a moving interface \citep{harvie2006analysis}. During the impact, the drop interface undergoes significant deceleration as it reaches its maximum spreading diameter. To mitigate the influence of spurious flows at this critical point, an additional criterion developed by \cite{deshpande2012evaluating} is incorporated. This criterion is dependent on fluid properties and can be expressed as follows:
\begin{equation*}
    \label{IJMF2023:eqn:dT4ST}
    \Delta t \leq \max \left( 10 \frac{\mu h}{\sigma},0.1 \sqrt{\frac{\rho h^3}{\sigma}}\right).
\end{equation*}

The surface tension at the liquid-air interface depends on the amount of surfactants present. The interface surfactant concentration can be modified by changing the interface surface area or via a transfer process involving the bulk or the wall. To prevent abrupt fluctuations in surface tension arising from the transfer of surfactants, we limit the source-sink terms using
\begin{equation*}
    \label{IJMF2023:eqn:dT4Sorption}
    \Delta t \leq C \max\left( \frac{\Gamma_{\textrm{m}}}{j_{\textrm{IB}} h}  \right).
\end{equation*}


\subsection{Kistler model} \label{IJMF2023:sec:Kistler}
Numerous models are available for the numerical treatment of the contact angle \citep{malgarinos2014vof,esteban2023contact}.
In this study, we have implemented the Kistler model \citep{kistler1993hydrodynamics} to the OpenFOAM software. The Kistler model calculates contact angles depending on the contact line velocity $u_\textrm{CL}$ (positive when spreading and negative when receding) and their values can surpass the limits of static advancing and receding contact angles, denoted as $\theta_\textrm{A}$ and $\theta_\textrm{R}$, respectively. The contact angle $\theta$ calculated with Kistler model uses the Hoffman function $f_\textrm{Hoff}$ as follows:
\begin{equation}
\label{IJMF2023:eqn:KislerCA}
\theta = f_\textrm{Hoff} \left( \frac{\mu \, u_\textrm{CL}}{\sigma} + f_\textrm{Hoff}^{-1} \left( \theta_\textrm{A$\vert$R}\right) \right),
\end{equation}
where $\theta_\textrm{A$\vert$R}$ represents the static contact angle during the spreading phase ($\theta_\textrm{A}$) and the receding phase ($\theta_\textrm{R}$). If these specific values are not available, an approximate value of $\theta_0$ is used instead. The Hoffman function $f_\textrm{Hoff}$ is defined as
\begin{equation}
\label{IJMF2023:eqn:Hoffman}
f_\textrm{Hoff}(\theta) = \arccos{\left(1-2 \tanh{\left(5.16 \left( \frac{\theta}{1+1.31 \theta^{0.99}} \right)^{0.706}\right)}\right)},
\end{equation}
where $\theta$ is the input contact angle.

In our investigation of surfactant-laden drop impactions, the Kistler model (\ref{IJMF2023:eqn:KislerCA}) is particularly relevant for two primary reasons. Firstly, the model is capable of describing contact angles that exceed the static advancing and receding contact angles. Secondly, the model explicitly incorporates surface tension in its formulation.

\subsection{Adaptive mesh}  \label{IJMF2023:sec:adaptMesh}

OpenFOAM provides a built-in mesh generator called blockMesh, enabling the creation of meshes that can be directly imported into the software. In addition, OpenFOAM incorporates an adaptive mesh function for three-dimensional (3D) domains consisting of hexahedrons. This function divides cells requiring refinement into four sub-cells by halving each dimension. However, OpenFOAM does not support adaptive meshing for wedge domains. To address this limitation, we implemented the code from \citep{rettenmaier2019load} that allows for the refinement of both hexahedral cells and prism cells.



\bibliographystyle{elsarticle-harv}\biboptions{authoryear}
\bibliography{combined}

\begin{thebibliography}{60}
\expandafter\ifx\csname natexlab\endcsname\relax\def\natexlab#1{#1}\fi
\providecommand{\url}[1]{\texttt{#1}}
\providecommand{\href}[2]{#2}
\providecommand{\path}[1]{#1}
\providecommand{\DOIprefix}{doi:}
\providecommand{\ArXivprefix}{arXiv:}
\providecommand{\URLprefix}{URL: }
\providecommand{\Pubmedprefix}{pmid:}
\providecommand{\doi}[1]{\href{http://dx.doi.org/#1}{\path{#1}}}
\providecommand{\Pubmed}[1]{\href{pmid:#1}{\path{#1}}}
\providecommand{\bibinfo}[2]{#2}
\ifx\xfnm\relax \def\xfnm[#1]{\unskip,\space#1}\fi
\bibitem[{Antritter et~al.(2019)Antritter, Hachmann, Gambaryan-Roisman, Buck
  and Stephan}]{antritter2019spreading}
\bibinfo{author}{Antritter, T.}, \bibinfo{author}{Hachmann, P.},
  \bibinfo{author}{Gambaryan-Roisman, T.}, \bibinfo{author}{Buck, B.},
  \bibinfo{author}{Stephan, P.}, \bibinfo{year}{2019}.
\newblock \bibinfo{title}{Spreading of micrometer-sized droplets under the
  influence of insoluble and soluble surfactants: A numerical study}.
\newblock \bibinfo{journal}{Colloids and Interfaces} \bibinfo{volume}{3},
  \bibinfo{pages}{56}.
\bibitem[{Badra et~al.(2018)Badra, Zahaf, Alla and
  Roques-Carmes}]{badra2018numerical}
\bibinfo{author}{Badra, A.T.}, \bibinfo{author}{Zahaf, H.},
  \bibinfo{author}{Alla, H.}, \bibinfo{author}{Roques-Carmes, T.},
  \bibinfo{year}{2018}.
\newblock \bibinfo{title}{A numerical model of superspreading surfactants on
  hydrophobic surface}.
\newblock \bibinfo{journal}{Physics of Fluids} \bibinfo{volume}{30},
  \bibinfo{pages}{092102}.
\bibitem[{Bardi and Osher(1991)}]{bardi1991nonconvex}
\bibinfo{author}{Bardi, M.}, \bibinfo{author}{Osher, S.}, \bibinfo{year}{1991}.
\newblock \bibinfo{title}{The nonconvex multidimensional riemann problem for
  hamilton--jacobi equations}.
\newblock \bibinfo{journal}{SIAM Journal on Mathematical Analysis}
  \bibinfo{volume}{22}, \bibinfo{pages}{344--351}.
\bibitem[{Bera et~al.(2016)Bera, Duits, Stuart, Van Den~Ende and
  Mugele}]{bera2016surfactant}
\bibinfo{author}{Bera, B.}, \bibinfo{author}{Duits, M.H.},
  \bibinfo{author}{Stuart, M.C.}, \bibinfo{author}{Van Den~Ende, D.},
  \bibinfo{author}{Mugele, F.}, \bibinfo{year}{2016}.
\newblock \bibinfo{title}{Surfactant induced autophobing}.
\newblock \bibinfo{journal}{Soft Matter} \bibinfo{volume}{12},
  \bibinfo{pages}{4562--4571}.
\bibitem[{Bonn et~al.(2009)Bonn, Eggers, Indekeu, Meunier and
  Rolley}]{bonn2009wetting}
\bibinfo{author}{Bonn, D.}, \bibinfo{author}{Eggers, J.},
  \bibinfo{author}{Indekeu, J.}, \bibinfo{author}{Meunier, J.},
  \bibinfo{author}{Rolley, E.}, \bibinfo{year}{2009}.
\newblock \bibinfo{title}{Wetting and spreading}.
\newblock \bibinfo{journal}{Reviews of Modern Physics} \bibinfo{volume}{81},
  \bibinfo{pages}{739}.
\bibitem[{Broom and Willmott(2023)}]{broom2023water}
\bibinfo{author}{Broom, M.}, \bibinfo{author}{Willmott, G.R.},
  \bibinfo{year}{2023}.
\newblock \bibinfo{title}{Water drop impacts on regular micropillar arrays:
  Asymmetric spreading}.
\newblock \bibinfo{journal}{Physics of Fluids} \bibinfo{volume}{35}.
\bibitem[{Bussmann et~al.(2000)Bussmann, Chandra and
  Mostaghimi}]{bussmann2000modeling}
\bibinfo{author}{Bussmann, M.}, \bibinfo{author}{Chandra, S.},
  \bibinfo{author}{Mostaghimi, J.}, \bibinfo{year}{2000}.
\newblock \bibinfo{title}{Modeling the splash of a droplet impacting a solid
  surface}.
\newblock \bibinfo{journal}{Physics of Fluids} \bibinfo{volume}{12},
  \bibinfo{pages}{3121--3132}.
\bibitem[{Chang and Franses(1992)}]{chang1992modified}
\bibinfo{author}{Chang, C.H.}, \bibinfo{author}{Franses, E.},
  \bibinfo{year}{1992}.
\newblock \bibinfo{title}{Modified langmuir—hinselwood kinetics for dynamic
  adsorption of surfactants at the air/water interface}.
\newblock \bibinfo{journal}{Colloids and Surfaces} \bibinfo{volume}{69},
  \bibinfo{pages}{189--201}.
\bibitem[{Cimpeanu and Papageorgiou(2018)}]{cimpeanu2018three}
\bibinfo{author}{Cimpeanu, R.}, \bibinfo{author}{Papageorgiou, D.T.},
  \bibinfo{year}{2018}.
\newblock \bibinfo{title}{Three-dimensional high speed drop impact onto solid
  surfaces at arbitrary angles}.
\newblock \bibinfo{journal}{International Journal of Multiphase Flow}
  \bibinfo{volume}{107}, \bibinfo{pages}{192--207}.
\bibitem[{Cooper-White et~al.(2002)Cooper-White, Crooks, Chockalingam and
  Boger}]{cooper2002dynamics}
\bibinfo{author}{Cooper-White, J.J.}, \bibinfo{author}{Crooks, R.C.},
  \bibinfo{author}{Chockalingam, K.}, \bibinfo{author}{Boger, D.V.},
  \bibinfo{year}{2002}.
\newblock \bibinfo{title}{Dynamics of polymer-surfactant complexes:
  Elongational properties and drop impact behavior}.
\newblock \bibinfo{journal}{Industrial \& Engineering Chemistry Research}
  \bibinfo{volume}{41}, \bibinfo{pages}{6443--6459}.
\bibitem[{Craster and Matar(2007)}]{craster2007autophobing}
\bibinfo{author}{Craster, R.}, \bibinfo{author}{Matar, O.},
  \bibinfo{year}{2007}.
\newblock \bibinfo{title}{On autophobing in surfactant-driven thin films}.
\newblock \bibinfo{journal}{Langmuir} \bibinfo{volume}{23},
  \bibinfo{pages}{2588--2601}.
\bibitem[{Crooks et~al.(2001)Crooks, Cooper-White and Boger}]{crooks2001role}
\bibinfo{author}{Crooks, R.}, \bibinfo{author}{Cooper-White, J.},
  \bibinfo{author}{Boger, D.V.}, \bibinfo{year}{2001}.
\newblock \bibinfo{title}{The role of dynamic surface tension and elasticity on
  the dynamics of drop impact}.
\newblock \bibinfo{journal}{Chemical Engineering Science} \bibinfo{volume}{56},
  \bibinfo{pages}{5575--5592}.
\bibitem[{Danov et~al.(2006)Danov, Kralchevsky, Denkov, Ananthapadmanabhan and
  Lips}]{danov2006mass}
\bibinfo{author}{Danov, K.}, \bibinfo{author}{Kralchevsky, P.},
  \bibinfo{author}{Denkov, N.}, \bibinfo{author}{Ananthapadmanabhan, K.},
  \bibinfo{author}{Lips, A.}, \bibinfo{year}{2006}.
\newblock \bibinfo{title}{Mass transport in micellar surfactant solutions: 2.
  theoretical modeling of adsorption at a quiescent interface}.
\newblock \bibinfo{journal}{Advances in Colloid and Interface Science}
  \bibinfo{volume}{119}, \bibinfo{pages}{17--33}.
\bibitem[{Debnath et~al.(2023)Debnath, Verma, Kumar and
  Balakrishnan}]{debnath2023understanding}
\bibinfo{author}{Debnath, D.}, \bibinfo{author}{Verma, D.},
  \bibinfo{author}{Kumar, P.}, \bibinfo{author}{Balakrishnan, V.},
  \bibinfo{year}{2023}.
\newblock \bibinfo{title}{Understanding the impact dynamics of droplets on
  superhydrophobic surface}.
\newblock \bibinfo{journal}{International Journal of Multiphase Flow}
  \bibinfo{volume}{159}, \bibinfo{pages}{104344}.
\bibitem[{Derby(2010)}]{derby2010inkjet}
\bibinfo{author}{Derby, B.}, \bibinfo{year}{2010}.
\newblock \bibinfo{title}{Inkjet printing of functional and structural
  materials: fluid property requirements, feature stability, and resolution}.
\newblock \bibinfo{journal}{Annual Review of Materials Research}
  \bibinfo{volume}{40}, \bibinfo{pages}{395--414}.
\bibitem[{Deshpande et~al.(2012)Deshpande, Anumolu and
  Trujillo}]{deshpande2012evaluating}
\bibinfo{author}{Deshpande, S.S.}, \bibinfo{author}{Anumolu, L.},
  \bibinfo{author}{Trujillo, M.F.}, \bibinfo{year}{2012}.
\newblock \bibinfo{title}{Evaluating the performance of the two-phase flow
  solver interfoam}.
\newblock \bibinfo{journal}{Computational Science \& Discovery}
  \bibinfo{volume}{5}, \bibinfo{pages}{014016}.
\bibitem[{Dorr et~al.(2016)Dorr, Forster, Mayo, McCue, Kempthorne, Hanan,
  Turner, Belward, Young and Zabkiewicz}]{dorr2016spray}
\bibinfo{author}{Dorr, G.J.}, \bibinfo{author}{Forster, W.A.},
  \bibinfo{author}{Mayo, L.C.}, \bibinfo{author}{McCue, S.W.},
  \bibinfo{author}{Kempthorne, D.M.}, \bibinfo{author}{Hanan, J.},
  \bibinfo{author}{Turner, I.W.}, \bibinfo{author}{Belward, J.A.},
  \bibinfo{author}{Young, J.}, \bibinfo{author}{Zabkiewicz, J.A.},
  \bibinfo{year}{2016}.
\newblock \bibinfo{title}{Spray retention on whole plants: modelling,
  simulations and experiments}.
\newblock \bibinfo{journal}{Crop Protection} \bibinfo{volume}{88},
  \bibinfo{pages}{118--130}.
\bibitem[{Dorr et~al.(2014)Dorr, Kempthorne, Mayo, Forster, Zabkiewicz, McCue,
  Belward, Turner and Hanan}]{dorr2014towards}
\bibinfo{author}{Dorr, G.J.}, \bibinfo{author}{Kempthorne, D.M.},
  \bibinfo{author}{Mayo, L.C.}, \bibinfo{author}{Forster, W.A.},
  \bibinfo{author}{Zabkiewicz, J.A.}, \bibinfo{author}{McCue, S.W.},
  \bibinfo{author}{Belward, J.A.}, \bibinfo{author}{Turner, I.W.},
  \bibinfo{author}{Hanan, J.}, \bibinfo{year}{2014}.
\newblock \bibinfo{title}{Towards a model of spray--canopy interactions:
  interception, shatter, bounce and retention of droplets on horizontal
  leaves}.
\newblock \bibinfo{journal}{Ecological Modelling} \bibinfo{volume}{290},
  \bibinfo{pages}{94--101}.
\bibitem[{Dorr et~al.(2015)Dorr, Wang, Mayo, McCue, Forster, Hanan and
  He}]{dorr2015impaction}
\bibinfo{author}{Dorr, G.J.}, \bibinfo{author}{Wang, S.},
  \bibinfo{author}{Mayo, L.C.}, \bibinfo{author}{McCue, S.W.},
  \bibinfo{author}{Forster, W.A.}, \bibinfo{author}{Hanan, J.},
  \bibinfo{author}{He, X.}, \bibinfo{year}{2015}.
\newblock \bibinfo{title}{Impaction of spray droplets on leaves: influence of
  formulation and leaf character on shatter, bounce and adhesion}.
\newblock \bibinfo{journal}{Experiments in Fluids} \bibinfo{volume}{56},
  \bibinfo{pages}{1--17}.
\bibitem[{Esmaeili et~al.(2021)Esmaeili, Mir and
  Mohammadi}]{esmaeili2021further}
\bibinfo{author}{Esmaeili, A.R.}, \bibinfo{author}{Mir, N.},
  \bibinfo{author}{Mohammadi, R.}, \bibinfo{year}{2021}.
\newblock \bibinfo{title}{Further step toward a comprehensive understanding of
  the effect of surfactant additions on altering the impact dynamics of water
  droplets}.
\newblock \bibinfo{journal}{Langmuir} \bibinfo{volume}{37},
  \bibinfo{pages}{841--851}.
\bibitem[{Esteban et~al.(2023)Esteban, G{\'o}mez, Zanzi, L{\'o}pez, Bussmann
  and Hern{\'a}ndez}]{esteban2023contact}
\bibinfo{author}{Esteban, A.}, \bibinfo{author}{G{\'o}mez, P.},
  \bibinfo{author}{Zanzi, C.}, \bibinfo{author}{L{\'o}pez, J.},
  \bibinfo{author}{Bussmann, M.}, \bibinfo{author}{Hern{\'a}ndez, J.},
  \bibinfo{year}{2023}.
\newblock \bibinfo{title}{A contact line force model for the simulation of drop
  impacts on solid surfaces using volume of fluid methods}.
\newblock \bibinfo{journal}{Computers \& Fluids} , \bibinfo{pages}{105946}.
\bibitem[{Fainerman et~al.(2006)Fainerman, Mys, Makievski, Petkov and
  Miller}]{fainerman2006dynamic}
\bibinfo{author}{Fainerman, V.}, \bibinfo{author}{Mys, V.},
  \bibinfo{author}{Makievski, A.}, \bibinfo{author}{Petkov, J.},
  \bibinfo{author}{Miller, R.}, \bibinfo{year}{2006}.
\newblock \bibinfo{title}{Dynamic surface tension of micellar solutions in the
  millisecond and submillisecond time range}.
\newblock \bibinfo{journal}{Journal of Colloid and Interface Science}
  \bibinfo{volume}{302}, \bibinfo{pages}{40--46}.
\bibitem[{Fang et~al.(2022)Fang, Zhang, Jiang, Lv, Sun, Li, Song and
  Feng}]{fang2022drop}
\bibinfo{author}{Fang, W.}, \bibinfo{author}{Zhang, K.},
  \bibinfo{author}{Jiang, Q.}, \bibinfo{author}{Lv, C.}, \bibinfo{author}{Sun,
  C.}, \bibinfo{author}{Li, Q.}, \bibinfo{author}{Song, Y.},
  \bibinfo{author}{Feng, X.Q.}, \bibinfo{year}{2022}.
\newblock \bibinfo{title}{Drop impact dynamics on solid surfaces}.
\newblock \bibinfo{journal}{Applied Physics Letters} \bibinfo{volume}{121}.
\bibitem[{Ganesan(2015)}]{ganesan2015simulations}
\bibinfo{author}{Ganesan, S.}, \bibinfo{year}{2015}.
\newblock \bibinfo{title}{Simulations of impinging droplets with
  surfactant-dependent dynamic contact angle}.
\newblock \bibinfo{journal}{Journal of Computational Physics}
  \bibinfo{volume}{301}, \bibinfo{pages}{178--200}.
\bibitem[{Gao and Liu(2021)}]{gao2021surfactant}
\bibinfo{author}{Gao, Y.}, \bibinfo{author}{Liu, J.G.}, \bibinfo{year}{2021}.
\newblock \bibinfo{title}{Surfactant-dependent contact line dynamics and
  droplet spreading on textured substrates: Derivations and computations}.
\newblock \bibinfo{journal}{Physica D: Nonlinear Phenomena}
  \bibinfo{volume}{428}, \bibinfo{pages}{133067}.
\bibitem[{Gao et~al.(2020)Gao, Lu, Zhang, Shi, Li, Zhao, Liu, Yang, Du and
  Fan}]{gao2020wetting}
\bibinfo{author}{Gao, Y.}, \bibinfo{author}{Lu, J.}, \bibinfo{author}{Zhang,
  P.}, \bibinfo{author}{Shi, G.}, \bibinfo{author}{Li, Y.},
  \bibinfo{author}{Zhao, J.}, \bibinfo{author}{Liu, Z.}, \bibinfo{author}{Yang,
  J.}, \bibinfo{author}{Du, F.}, \bibinfo{author}{Fan, R.},
  \bibinfo{year}{2020}.
\newblock \bibinfo{title}{Wetting and adhesion behavior on apple tree leaf
  surface by adding different surfactants}.
\newblock \bibinfo{journal}{Colloids and Surfaces B: Biointerfaces}
  \bibinfo{volume}{187}, \bibinfo{pages}{110602}.
\bibitem[{Gopala and Van~Wachem(2008)}]{gopala2008volume}
\bibinfo{author}{Gopala, V.R.}, \bibinfo{author}{Van~Wachem, B.G.},
  \bibinfo{year}{2008}.
\newblock \bibinfo{title}{Volume of fluid methods for immiscible-fluid and
  free-surface flows}.
\newblock \bibinfo{journal}{Chemical Engineering Journal}
  \bibinfo{volume}{141}, \bibinfo{pages}{204--221}.
\bibitem[{Grishaev et~al.(2015)Grishaev, Iorio, Dubois and
  Amirfazli}]{grishaev2015complex}
\bibinfo{author}{Grishaev, V.}, \bibinfo{author}{Iorio, C.S.},
  \bibinfo{author}{Dubois, F.}, \bibinfo{author}{Amirfazli, A.},
  \bibinfo{year}{2015}.
\newblock \bibinfo{title}{Complex drop impact morphology}.
\newblock \bibinfo{journal}{Langmuir} \bibinfo{volume}{31},
  \bibinfo{pages}{9833--9844}.
\bibitem[{Gu et~al.(2019)Gu, Wen, Yao and Yu}]{gu2019volume}
\bibinfo{author}{Gu, Z.}, \bibinfo{author}{Wen, H.}, \bibinfo{author}{Yao, Y.},
  \bibinfo{author}{Yu, C.}, \bibinfo{year}{2019}.
\newblock \bibinfo{title}{A volume of fluid method algorithm for simulation of
  surface tension dominant two-phase flows}.
\newblock \bibinfo{journal}{Numerical Heat Transfer, Part B: Fundamentals}
  \bibinfo{volume}{76}, \bibinfo{pages}{1--17}.
\bibitem[{Gueyffier et~al.(1999)Gueyffier, Li, Nadim, Scardovelli and
  Zaleski}]{gueyffier1999volume}
\bibinfo{author}{Gueyffier, D.}, \bibinfo{author}{Li, J.},
  \bibinfo{author}{Nadim, A.}, \bibinfo{author}{Scardovelli, R.},
  \bibinfo{author}{Zaleski, S.}, \bibinfo{year}{1999}.
\newblock \bibinfo{title}{Volume-of-fluid interface tracking with smoothed
  surface stress methods for three-dimensional flows}.
\newblock \bibinfo{journal}{Journal of Computational Physics}
  \bibinfo{volume}{152}, \bibinfo{pages}{423--456}.
\bibitem[{Haghshenas et~al.(2017)Haghshenas, Wilson and
  Kumar}]{haghshenas2017algebraic}
\bibinfo{author}{Haghshenas, M.}, \bibinfo{author}{Wilson, J.A.},
  \bibinfo{author}{Kumar, R.}, \bibinfo{year}{2017}.
\newblock \bibinfo{title}{Algebraic coupled level set-volume of fluid method
  for surface tension dominant two-phase flows}.
\newblock \bibinfo{journal}{International Journal of Multiphase Flow}
  \bibinfo{volume}{90}, \bibinfo{pages}{13--28}.
\bibitem[{Harvie et~al.(2006)Harvie, Davidson and Rudman}]{harvie2006analysis}
\bibinfo{author}{Harvie, D.J.}, \bibinfo{author}{Davidson, M.},
  \bibinfo{author}{Rudman, M.}, \bibinfo{year}{2006}.
\newblock \bibinfo{title}{An analysis of parasitic current generation in volume
  of fluid simulations}.
\newblock \bibinfo{journal}{Applied Mathematical Modelling}
  \bibinfo{volume}{30}, \bibinfo{pages}{1056--1066}.
\bibitem[{Henman et~al.(2023)Henman, Smith and
  Tiwari}]{henman2023computational}
\bibinfo{author}{Henman, N.I.}, \bibinfo{author}{Smith, F.T.},
  \bibinfo{author}{Tiwari, M.K.}, \bibinfo{year}{2023}.
\newblock \bibinfo{title}{Computational study of early-time droplet impact
  dynamics on textured and lubricant-infused surfaces}.
\newblock \bibinfo{journal}{International Journal of Multiphase Flow}
  \bibinfo{volume}{161}, \bibinfo{pages}{104398}.
\bibitem[{Hu and Liu(2022)}]{hu20223d}
\bibinfo{author}{Hu, A.}, \bibinfo{author}{Liu, D.}, \bibinfo{year}{2022}.
\newblock \bibinfo{title}{3d simulation of micro droplet impact on the
  structured superhydrophobic surface}.
\newblock \bibinfo{journal}{International Journal of Multiphase Flow}
  \bibinfo{volume}{147}, \bibinfo{pages}{103887}.
\bibitem[{Huet et~al.(2020)Huet, Massinon, De~Cock, Forster, Zabkiewicz,
  Pethiyagoda, Moroney, Lebeau and McCue}]{huet2020image}
\bibinfo{author}{Huet, O.D.}, \bibinfo{author}{Massinon, M.},
  \bibinfo{author}{De~Cock, N.}, \bibinfo{author}{Forster, W.A.},
  \bibinfo{author}{Zabkiewicz, J.A.}, \bibinfo{author}{Pethiyagoda, R.},
  \bibinfo{author}{Moroney, T.J.}, \bibinfo{author}{Lebeau, F.},
  \bibinfo{author}{McCue, S.W.}, \bibinfo{year}{2020}.
\newblock \bibinfo{title}{Image analysis of shatter and pinning events on
  hard-to-wet leaf surfaces by drops containing surfactant}.
\newblock \bibinfo{journal}{Pest Management Science} \bibinfo{volume}{76},
  \bibinfo{pages}{3477--3486}.
\bibitem[{Josserand and Thoroddsen(2016)}]{josserand2016drop}
\bibinfo{author}{Josserand, C.}, \bibinfo{author}{Thoroddsen, S.T.},
  \bibinfo{year}{2016}.
\newblock \bibinfo{title}{Drop impact on a solid surface}.
\newblock \bibinfo{journal}{Annual Review of Fluid Mechanics}
  \bibinfo{volume}{48}, \bibinfo{pages}{365--391}.
\bibitem[{Karapetsas et~al.(2011)Karapetsas, Craster and
  Matar}]{karapetsas2011surfactant}
\bibinfo{author}{Karapetsas, G.}, \bibinfo{author}{Craster, R.V.},
  \bibinfo{author}{Matar, O.K.}, \bibinfo{year}{2011}.
\newblock \bibinfo{title}{On surfactant-enhanced spreading and superspreading
  of liquid drops on solid surfaces}.
\newblock \bibinfo{journal}{Journal of Fluid Mechanics} \bibinfo{volume}{670},
  \bibinfo{pages}{5--37}.
\bibitem[{Khojasteh et~al.(2016)Khojasteh, Kazerooni, Salarian and
  Kamali}]{khojasteh2016droplet}
\bibinfo{author}{Khojasteh, D.}, \bibinfo{author}{Kazerooni, M.},
  \bibinfo{author}{Salarian, S.}, \bibinfo{author}{Kamali, R.},
  \bibinfo{year}{2016}.
\newblock \bibinfo{title}{Droplet impact on superhydrophobic surfaces: A review
  of recent developments}.
\newblock \bibinfo{journal}{Journal of Industrial and Engineering Chemistry}
  \bibinfo{volume}{42}, \bibinfo{pages}{1--14}.
\bibitem[{Kistler(1993)}]{kistler1993hydrodynamics}
\bibinfo{author}{Kistler, S.F.}, \bibinfo{year}{1993}.
\newblock \bibinfo{title}{Hydrodynamics of wetting}.
\newblock \bibinfo{journal}{Wettability} \bibinfo{volume}{6},
  \bibinfo{pages}{311--430}.
\bibitem[{Landoll(1982)}]{landoll1982nonionic}
\bibinfo{author}{Landoll, L.}, \bibinfo{year}{1982}.
\newblock \bibinfo{title}{Nonionic polymer surfactants}.
\newblock \bibinfo{journal}{Journal of Polymer Science: Polymer Chemistry
  Edition} \bibinfo{volume}{20}, \bibinfo{pages}{443--455}.
\bibitem[{Malgarinos et~al.(2014)Malgarinos, Nikolopoulos, Marengo, Antonini
  and Gavaises}]{malgarinos2014vof}
\bibinfo{author}{Malgarinos, I.}, \bibinfo{author}{Nikolopoulos, N.},
  \bibinfo{author}{Marengo, M.}, \bibinfo{author}{Antonini, C.},
  \bibinfo{author}{Gavaises, M.}, \bibinfo{year}{2014}.
\newblock \bibinfo{title}{Vof simulations of the contact angle dynamics during
  the drop spreading: Standard models and a new wetting force model}.
\newblock \bibinfo{journal}{Advances in Colloid and Interface Science}
  \bibinfo{volume}{212}, \bibinfo{pages}{1--20}.
\bibitem[{Manikantan and Squires(2020)}]{manikantan2020surfactant}
\bibinfo{author}{Manikantan, H.}, \bibinfo{author}{Squires, T.M.},
  \bibinfo{year}{2020}.
\newblock \bibinfo{title}{Surfactant dynamics: hidden variables controlling
  fluid flows}.
\newblock \bibinfo{journal}{Journal of Fluid Mechanics} \bibinfo{volume}{892},
  \bibinfo{pages}{P1}.
\bibitem[{Massinon et~al.(2017)Massinon, De~Cock, Forster, Nairn, McCue,
  Zabkiewicz and Lebeau}]{massinon2017spray}
\bibinfo{author}{Massinon, M.}, \bibinfo{author}{De~Cock, N.},
  \bibinfo{author}{Forster, W.A.}, \bibinfo{author}{Nairn, J.J.},
  \bibinfo{author}{McCue, S.W.}, \bibinfo{author}{Zabkiewicz, J.A.},
  \bibinfo{author}{Lebeau, F.}, \bibinfo{year}{2017}.
\newblock \bibinfo{title}{Spray droplet impaction outcomes for different plant
  species and spray formulations}.
\newblock \bibinfo{journal}{Crop Protection} \bibinfo{volume}{99},
  \bibinfo{pages}{65--75}.
\bibitem[{Mehmani(2018)}]{mehmani2018wrinkle}
\bibinfo{author}{Mehmani, Y.}, \bibinfo{year}{2018}.
\newblock \bibinfo{title}{Wrinkle-free interface compression for two-fluid
  flows}.
\newblock \bibinfo{journal}{arXiv preprint arXiv:1811.09744} .
\bibitem[{Mousavi et~al.(2016)Mousavi, Kummer, Oberlack and
  Pelz}]{mousavi2016level}
\bibinfo{author}{Mousavi, R.}, \bibinfo{author}{Kummer, F.},
  \bibinfo{author}{Oberlack, M.}, \bibinfo{author}{Pelz, P.F.},
  \bibinfo{year}{2016}.
\newblock \bibinfo{title}{Level set method for simulating the interface
  kinematics: application of a discontinuous galerkin method}, in:
  \bibinfo{booktitle}{Proceedings of ECCOMAS congress}.
\bibitem[{Okagaki et~al.(2021)Okagaki, Yonomoto, Ishigaki and
  Hirose}]{okagaki2021numerical}
\bibinfo{author}{Okagaki, Y.}, \bibinfo{author}{Yonomoto, T.},
  \bibinfo{author}{Ishigaki, M.}, \bibinfo{author}{Hirose, Y.},
  \bibinfo{year}{2021}.
\newblock \bibinfo{title}{Numerical study on an interface compression method
  for the volume of fluid approach}.
\newblock \bibinfo{journal}{Fluids} \bibinfo{volume}{6}, \bibinfo{pages}{80}.
\bibitem[{{OpenCFD Ltd}(2004)}]{OpenFOAM}
\bibinfo{author}{{OpenCFD Ltd}}, \bibinfo{year}{2004}.
\newblock \bibinfo{title}{{OpenFOAM}}.
\newblock \bibinfo{howpublished}{\url{https://www.openfoam.com}}.
\newblock \bibinfo{note}{Accessed: 05-06-2023}.
\bibitem[{Pasandideh-Fard et~al.(1996)Pasandideh-Fard, Qiao, Chandra and
  Mostaghimi}]{pasandideh1996capillary}
\bibinfo{author}{Pasandideh-Fard, M.}, \bibinfo{author}{Qiao, Y.},
  \bibinfo{author}{Chandra, S.}, \bibinfo{author}{Mostaghimi, J.},
  \bibinfo{year}{1996}.
\newblock \bibinfo{title}{Capillary effects during droplet impact on a solid
  surface}.
\newblock \bibinfo{journal}{Physics of Fluids} \bibinfo{volume}{8},
  \bibinfo{pages}{650--659}.
\bibitem[{Peng et~al.(1999)Peng, Merriman, Osher, Zhao and Kang}]{peng1999pde}
\bibinfo{author}{Peng, D.}, \bibinfo{author}{Merriman, B.},
  \bibinfo{author}{Osher, S.}, \bibinfo{author}{Zhao, H.},
  \bibinfo{author}{Kang, M.}, \bibinfo{year}{1999}.
\newblock \bibinfo{title}{A pde-based fast local level set method}.
\newblock \bibinfo{journal}{Journal of Computational Physics}
  \bibinfo{volume}{155}, \bibinfo{pages}{410--438}.
\bibitem[{Rettenmaier et~al.(2019)Rettenmaier, Deising, Ouedraogo, Gjonaj,
  De~Gersem, Bothe, Tropea and Marschall}]{rettenmaier2019load}
\bibinfo{author}{Rettenmaier, D.}, \bibinfo{author}{Deising, D.},
  \bibinfo{author}{Ouedraogo, Y.}, \bibinfo{author}{Gjonaj, E.},
  \bibinfo{author}{De~Gersem, H.}, \bibinfo{author}{Bothe, D.},
  \bibinfo{author}{Tropea, C.}, \bibinfo{author}{Marschall, H.},
  \bibinfo{year}{2019}.
\newblock \bibinfo{title}{Load balanced 2d and 3d adaptive mesh refinement in
  openfoam}.
\newblock \bibinfo{journal}{SoftwareX} \bibinfo{volume}{10},
  \bibinfo{pages}{100317}.
\bibitem[{Rioboo et~al.(2002)Rioboo, Marengo and Tropea}]{rioboo2002time}
\bibinfo{author}{Rioboo, R.}, \bibinfo{author}{Marengo, M.},
  \bibinfo{author}{Tropea, C.}, \bibinfo{year}{2002}.
\newblock \bibinfo{title}{Time evolution of liquid drop impact onto solid, dry
  surfaces}.
\newblock \bibinfo{journal}{Experiments in Fluids} \bibinfo{volume}{33},
  \bibinfo{pages}{112--124}.
\bibitem[{Rioboo et~al.(2001)Rioboo, Tropea and Marengo}]{rioboo2001outcomes}
\bibinfo{author}{Rioboo, R.}, \bibinfo{author}{Tropea, C.},
  \bibinfo{author}{Marengo, M.}, \bibinfo{year}{2001}.
\newblock \bibinfo{title}{Outcomes from a drop impact on solid surfaces}.
\newblock \bibinfo{journal}{Atomization and Sprays} \bibinfo{volume}{11},
  \bibinfo{pages}{155--165}.
\bibitem[{Scardovelli and Zaleski(1999)}]{scardovelli1999direct}
\bibinfo{author}{Scardovelli, R.}, \bibinfo{author}{Zaleski, S.},
  \bibinfo{year}{1999}.
\newblock \bibinfo{title}{Direct numerical simulation of free-surface and
  interfacial flow}.
\newblock \bibinfo{journal}{Annual Review of Fluid Mechanics}
  \bibinfo{volume}{31}, \bibinfo{pages}{567--603}.
\bibitem[{Theodorakis et~al.(2015)Theodorakis, M{\"u}ller, Craster and
  Matar}]{theodorakis2015superspreading}
\bibinfo{author}{Theodorakis, P.E.}, \bibinfo{author}{M{\"u}ller, E.A.},
  \bibinfo{author}{Craster, R.V.}, \bibinfo{author}{Matar, O.K.},
  \bibinfo{year}{2015}.
\newblock \bibinfo{title}{Superspreading: Mechanisms and molecular design}.
\newblock \bibinfo{journal}{Langmuir} \bibinfo{volume}{31},
  \bibinfo{pages}{2304--2309}.
\bibitem[{Theodorakis et~al.(2019)Theodorakis, Smith, Craster, M{\"u}ller and
  Matar}]{theodorakis2019molecular}
\bibinfo{author}{Theodorakis, P.E.}, \bibinfo{author}{Smith, E.R.},
  \bibinfo{author}{Craster, R.V.}, \bibinfo{author}{M{\"u}ller, E.A.},
  \bibinfo{author}{Matar, O.K.}, \bibinfo{year}{2019}.
\newblock \bibinfo{title}{Molecular dynamics simulation of the superspreading
  of surfactant-laden droplets. a review}.
\newblock \bibinfo{journal}{Fluids} \bibinfo{volume}{4}, \bibinfo{pages}{176}.
\bibitem[{Varghese et~al.(2024)Varghese, Sykes, Quetzeri-Santiago,
  Castrej{\'o}n-Pita and Castrej{\'o}n-Pita}]{varghese2024effect}
\bibinfo{author}{Varghese, N.}, \bibinfo{author}{Sykes, T.C.},
  \bibinfo{author}{Quetzeri-Santiago, M.A.},
  \bibinfo{author}{Castrej{\'o}n-Pita, A.A.},
  \bibinfo{author}{Castrej{\'o}n-Pita, J.R.}, \bibinfo{year}{2024}.
\newblock \bibinfo{title}{Effect of surfactants on the splashing dynamics of
  drops impacting smooth substrates}.
\newblock \bibinfo{journal}{Langmuir} , \bibinfo{pages}{published online 5
  March 2024}.
\bibitem[{Wang et~al.(2022a)Wang, Guo and Zhang}]{wang2022unconditionally}
\bibinfo{author}{Wang, C.}, \bibinfo{author}{Guo, Y.}, \bibinfo{author}{Zhang,
  Z.}, \bibinfo{year}{2022}a.
\newblock \bibinfo{title}{Unconditionally energy stable and bound-preserving
  schemes for phase-field surfactant model with moving contact lines}.
\newblock \bibinfo{journal}{Journal of Scientific Computing}
  \bibinfo{volume}{92}, \bibinfo{pages}{20}.
\bibitem[{Wang et~al.(2022b)Wang, Lai and Zhang}]{wang2022energy}
\bibinfo{author}{Wang, C.}, \bibinfo{author}{Lai, M.C.},
  \bibinfo{author}{Zhang, Z.}, \bibinfo{year}{2022}b.
\newblock \bibinfo{title}{Energy stable methods for phase-field simulation of
  droplet impact with surfactants}.
\newblock \bibinfo{journal}{arXiv preprint arXiv:2212.08845} .
\bibitem[{Yokoi(2014)}]{yokoi2014density}
\bibinfo{author}{Yokoi, K.}, \bibinfo{year}{2014}.
\newblock \bibinfo{title}{A density-scaled continuum surface force model within
  a balanced force formulation}.
\newblock \bibinfo{journal}{Journal of Computational Physics}
  \bibinfo{volume}{278}, \bibinfo{pages}{221--228}.
\bibitem[{Zhang and Basaran(1997)}]{zhang1997dynamic}
\bibinfo{author}{Zhang, X.}, \bibinfo{author}{Basaran, O.A.},
  \bibinfo{year}{1997}.
\newblock \bibinfo{title}{Dynamic surface tension effects in impact of a drop
  with a solid surface}.
\newblock \bibinfo{journal}{Journal of Colloid and Interface Science}
  \bibinfo{volume}{187}, \bibinfo{pages}{166--178}.

\end{thebibliography}

\end{document}